\documentclass[a4paper,11pt]{article}
\pdfoutput=1 

\usepackage{jheppub} 

\usepackage[T1]{fontenc} 
\usepackage[utf8x]{inputenc}
\usepackage{color}
\usepackage{genyoungtabtikz, stmaryrd, mathrsfs, dsfont}

\def \de{\partial}

\newcommand{\so}{\mathfrak{so}}

\newcommand{\gl}{\mathfrak{gl}}

\newcommand{\tr}{\mathfrak{t}}
\newcommand{\dd}{\mathrm{d}}

\newcommand{\D}{\mathcal{D}}
\newcommand{\DD}{\mathscr{D}}

\newcommand{\M}{\mathscr{M}}
\newcommand{\B}{\mathscr{B}}

\newcommand{\End}{\mathrm{End}}
\newcommand{\R}{\mathbb{R}}

\newcommand{\N}{\mathbb{N}}

\newcommand{\V}{\mathcal{V}}

\Yboxdim{7pt} 
\Ylinethick{.8pt} 
\newcommand\s{s-1}
\newcommand\ka{k}
\newcommand\su{s-2}

\newcommand\depth{t}

\definecolor{rougef}{rgb}{0.56,0,0}
\definecolor{vertf}{rgb}{0,0.5,0}
\definecolor{bleuf}{rgb}{0,0,0.8}

\newcommand{\Rem}[1]{\textcolor{rougef}{\textbf{[#1]}}}

\newlength\foo

\title{\boldmath The Schouten tensor as a connection 
in the unfolding of 3D conformal higher-spin fields}

\author[a,b]{Thomas Basile,}
\author[a]{Roberto Bonezzi}
\author[a,1]{and Nicolas Boulanger\note{Associate Researcher of the F.R.S.-FNRS (Belgium)}}

\affiliation[a]{Group of Mechanics and Gravitation, Physique th\'eorique et math\'ematique\\
University of Mons -- UMONS, 20 Place du Parc, 7000 Mons, Belgium}
\affiliation[b]{Laboratoire de Math\'ematiques et Physique Th\'eorique, Unit\'e Mixte de Recherche $7350$ du CNRS\\
F\'ed\'eration de Recherche $2964$ Denis Poisson, Universit\'e Fran\c{c}ois Rabelais, Parc de Grandmont\\
37200 Tours, France}

\emailAdd{thomas.basile@umons.ac.be, roberto.bonezzi@umons.ac.be, 
nicolas.boulanger@umons.ac.be}

\abstract{A first-order differential equation is provided for a 
one-form, spin-$s$ connection valued in the two-row, width-$(s-1)$ 
Young tableau of $GL(5)\,$. The connection is glued to a zero-form
identified with the spin-$s$ Cotton tensor. 
The usual zero-Cotton equation 
for a symmetric, conformal spin-$s$ tensor gauge field in 3D is the flatness condition 
for the sum of the $GL(5)$ spin-$s$ and background connections.
This presentation of the equations allows to reformulate in a compact way 
the cohomological problem studied in 
\href{https://arxiv.org/abs/1511.07389}{1511.07389}, 
featuring the spin-$s$ Schouten tensor.
We provide full computational details for spin 3 and 4 and 
present the general spin-$s$ case in a compact way.} 

\begin{document} 
\maketitle
\flushbottom

\section{Introduction}
\label{sec:intro}

Conformal higher-spin theory is a fascinating field 
of investigation in its own and in relation with 
studies of string theory in an unbroken, tensionless phase. 
As a result, it has received a lot of attention, as for example  
\cite{Fradkin:1985am, Siegel:1988gd, Pope:1989vj, Fradkin:1989xt, Metsaev:1995jp, Preitschopf:1998ei, Shaynkman:2001ip, Segal:2002gd, Shaynkman:2004vu, Metsaev:2007rw, Marnelius:2008er, Vasiliev:2009ck, Bandos:2011wi, Bekaert:2012vt}, 
and more recently 
\cite{Beccaria:2014jxa, Shaynkman:2014fqa, Barnich:2015tma, Nilsson:2015pua, Fernando:2015tiu, Henneaux:2015cda, Haehnel:2016mlb, Beccaria:2016syk, Grigoriev:2016bzl, Metsaev:2016rpa} 
to cite but a few works on the subject.
A reason for the importance it possesses
is due to the fact that it combines two 
unifying symmetries encountered in field theory, namely 
higher-spin gauge symmetry and conformal symmetry, 
the latter being promoted to Weyl invariance
when gravity is considered.
It was also shown in \cite{Henneaux:2015cda,Henneaux:2016zlu} 
that conformal invariance appears on fixed-time spacetime slices 
for Hamiltonian systems of high (and low) spin gauge fields  
enjoying off-shell electric-magnetic duality,
another symmetry of utmost importance in fundamental physics.
This adds an extra motivation for the study of conformal higher-spin theories.

It is particularly convenient to study (super)conformal 
higher-spin theory in 2+1 dimensions \cite{Pope:1989vj,Fradkin:1989xt} 
since the gauge sector can be described by a Chern--Simons action; see e.g. \cite{Bergshoeff:2009tb, Nilsson:2013tva, Nilsson:2015pua,
Henneaux:2016zlu, Kuzenko:2016qdw} 
for recent related works. 
Particularly powerful indeed is the Cartan formulation 
of gauge systems, where gauge fields are incorporated into Lie algebra valued
local $p$-forms on a base manifold and field equations are given 
by integrable constraints on generalised curvatures.
In the context of higher-spin theory, one refers to it as 
the \emph{unfolded} formulation of a gauge system \cite{Vasiliev:1992gr}, 
see also \cite{Shaynkman:2000ts,Skvortsov:2008vs,Boulanger:2008up}. 
In particular, the local propagating degrees of freedom are fully captured 
by an infinite set of zero-forms building up a representation 
of the isometry algebra of the background spacetime.

In \cite{Boulanger:2014vya},  
all the possible unfolded systems were classified 
for tensor gauge fields in 3D with (A)dS and flat backgrounds, 
at the level of the zero-form module; 
see \cite{Zinoviev:2015sra,Buchbinder:2016jgk} for related works including 
supersymmetric extensions. 
In particular, in \cite{Boulanger:2014vya} 
the zero-forms of conformal spin-s fields, starting with the 
Cotton tensor, were discussed 
from the massless limit of topologically massive gravity. 
One of the motivations underlying 
the present note is to provide an explicit 1-form module 
of the spacetime isometry algebra 
that glues to the Cotton zero-form module, thereby completing 
the discussion of \cite{Boulanger:2014vya} for 3D conformal fields. 

Of course, the procedure to build $p$-form module for 
conformal systems is well known, see in 
particular the classification of $\mathfrak{so}(2,d)$ modules 
provided in \cite{Shaynkman:2004vu} and used in \cite{Vasiliev:2009ck}
to which we refer for an extensive study of unfolding of conformal systems. 
In \cite{Shaynkman:2004vu}, the authors showed that the classification of
$\so(2,d)$ invariant systems of differential equations is equivalent
to the classification of $\so(2,d)$ modules\,\footnote{Actually, the 
results of \cite{Shaynkman:2004vu} hold for a whole class of semi-simple 
Lie algebra. For the sake of conciseness though, we will restrict the 
discussion here to the conformal algebra, of interest for this paper.}. 
The idea of their proof is based on the crucial role played by a 
nilpotent operator, $\sigma^-\,\,$, related  
to the background spacetime part of the connection. 
In the unfolded approach, the equations of motion generically
have the form $\DD \omega_{[p]} = 0\,$, where 
$\omega_{[p]} \in \Omega^p(\M) \otimes \D$ denotes a set of $p$-forms taking values in some $\mathbb{N}$-graded module 
$\D = \bigoplus_{n \in \N} \D_n\,$ of a Lie algebra $\mathfrak{g}\,$, 
and $\DD$ is a nilpotent operator. 
Taking as a special case conformal higher spin fields in flat background, 
to fix the ideas, this operator can be
decomposed in the form $\DD := \mathrm d + \sigma^-\,$, 
where $\mathrm d^2 =0\, ,\, ({\sigma^-}){}^2 = 0$ and 
$\left\{ \mathrm d, \sigma^- \right\} = 0\,$. 
The 1-form-valued operator $\sigma^-$ decreases the $\mathbb{N}$ degree 
by one unit. 
Within this setup, it turns out that  
all the quantities of physical interest are encapsulated in 
cohomology classes of $\sigma^-\,$. 
Denoting the background vielbeins by $h^a\,$, it appears  
--- see for example Section 4 of \cite{Boulanger:2008up} ---  
that $\sigma^-$ is related to the representation $\rho(P_a\vert h)$ 
of the translation generators $P_a$ on the module $\D$ by 
$\rho(P_a\vert h)= i \frac{\partial}{\partial h^a}\,\sigma^-\,$, 
with $\rho : \so(2,d) \rightarrow \End(\D)$ being the
representation. 
Indeed, the equation $\DD \omega_{[p]}=0$ can be viewed as a covariant
constancy condition. 

Let us sketch the central result of
\cite{Shaynkman:2004vu}. Consider the two trivial bundles $\B$ and $\B_c$ over
$\R^d$ whose fiber are respectively $\D$ and $\D_c\,$, 
the latter being a  submodule of
$\D$ composed of $\sigma^-\,$-closed $p\,$-forms. The authors of
\cite{Shaynkman:2004vu} proved that:
\begin{itemize}
\item For any section of $\B_c$ there exists a $\DD$ closed
  representative in $\B$ \it if and only if \rm $H^{p+1}(\tr, \D)$,
  the $p+1$ cohomology group of the translation subalgebra $\tr$ of the
  conformal group with coefficient in $\D$ is empty, i.e. 
  $H^{p+1}(\tr,\D)\cong 0\,$;
\item If $H^{p+1}(\tr, \D) \ncong 0$, there exists a system of
  differential equations such that any section of $\B_c$ is a solution
  of it \it if and only if \rm it admits a $\DD$ closed representative
  in $\B$.
\end{itemize}
Retrospectively, this result can be seen as the translation of the
fact that non-equivalent differential equations on dynamical
fields are in correspondence with the
$\sigma^-$-cohomology group at form degree $(p+1)\,$ 
on $\D\,$, if the dynamical fields sit at form degree $p$ on $\D\,$ 
\cite{Shaynkman:2000ts,Skvortsov:2008vs,Boulanger:2008up,Ponomarev:2010st}.
The advantage of this approach is that one can then relate this 
cohomology group to purely group theoretical data of $\D$: if $\D$ is an 
irreducible module such that it is obtained as a quotient of a 
generalized
Verma module $\V$ by its maximal irreducible submodule, then the cohomology group is 
isomorphic to a subspace of $\V\,$, 
determined by singular vectors. As a consequence,
the possible conformally invariant unfolded systems of equations are in 
correspondence with the possible irreducible $\so(2,d)$ modules, 
see \cite{Shaynkman:2004vu,Vasiliev:2009ck}.\\

In the present paper, therefore, we do not want to further discuss 
the $\sigma^-$-cohomologies for $\mathfrak{so}(2,d)$-invariant 
systems. It is well-known that the dynamical conformal spin-s 
field is represented by a symmetric, rank-s, \emph{traceless} tensor 
$\varphi_{\mu_1\ldots \mu_s}\,$ referred to as the Fradkin--Tseytlin 
spin-$s$ field. 
Instead, we aim at providing an off-shell conformal calculus where the gauge potential 
is \emph{traceful} and focus on the cases of maximally symmetric 
spacetimes\footnote{For recent extensions to more general, 
Bach-flat backgrounds, see e.g. \cite{Nutma:2014pua,Grigoriev:2016bzl}.} 
in $d = 3$ dimensions with metric of components $\bar{g}_{\mu\nu}\,$.
In other terms, the gauge symmetry 
$\delta_\sigma\varphi_{\mu_1\ldots \mu_s} = \frac{s(s-1)}{2}\,\bar{g}_{(\mu_1\mu_2}\,\sigma_{\mu_3\ldots\mu_s)}$ 
is still activated, together with the spin-$s$ linearised diffeomorphisms 
$\delta_\xi\varphi_{\mu_1\ldots \mu_s} = s\,\bar\nabla_{(\mu_1}\xi_{\mu_2\ldots\mu_s)}\,$.

As was stressed in the recent papers
\cite{Henneaux:2015cda,Henneaux:2016zlu}, an 
important role is played, in conformal spin-$s$ geometry, 
by the Schouten tensor, although it does not sit in any $\sigma^-$ cohomology class 
whatsoever. Still, it is an instrumental off-shell tensor in conformal geometry. 
See for example 
\cite{Francois:2015pga} and references therein for recent works highlighting 
the central role of the Schouten tensor in the Cartan normal 
connection of conformal gravity.
The relevance of the spin-$2$ Schouten tensor in 3D topologically massive
gravity was emphasised in \cite{Deser:1981wh}. 
In fact, as we exhibit in this paper, 
the discussion given in \cite{Henneaux:2015cda} 
on the Bianchi identities and gauge transformations for the spin-$s$ 
Schouten tensor can receive a Cartan-like, or frame-like, formulation
that highlights the underlying conformal geometry. 
More precisely, we will explicitly show where 
and how the 3D spin-$s$ Schouten tensor sits in the one-form module
describing conformal spin-$s$ theory in the frame-like formulation.   
As a by-product, we will show that the cohomological problem studied 
in \cite{Henneaux:2015cda} can receive a more geometrical 
formulation through the introduction of a conformal, flat connection,
and how a differential de Rham complex suffices for that matter.

With these very specific and concrete goals in mind 
and for computational convenience,  
we are led to extend the Lie algebra from 
$\mathfrak{so}(2,3)\,$, the conformal algebra in 3D,  
to the general linear $\mathfrak{gl}(5,\R)$ algebra.  
We do this by introducing contractible cycles in the one-form module of the 
conformal connection, following the general discussion presented 
in the Section 3.4 of \cite{Boulanger:2008up}. 
In turn, the $\mathfrak{gl}(5,\R)$-valued spin-$s$ connection 
is decomposed under $\mathfrak{gl}(3,\R)\,$. 
The extension from $\so(1,2)$ to $\mathfrak{gl}(3,\mathbb{R})$ 
proved already useful \cite{Boulanger:2014vya} 
in unfolding new massive spin-$s$ systems and extensions thereof and 
is, here as well, instrumental in order to achieve explicit, by-hand
computations involving Schouten and Cotton tensors for conformal 
spin-$s$ fields in 3D. \\

The layout of the paper is as follows,
\begin{itemize}
\item In \hyperref[sec:2]{Section \ref{sec:2}} we expose the general setting in which 
we unfold the conformal higher spin free equations of motion;
\item In \hyperref[sec:3]{Section \ref{sec:3}} and \hyperref[sec:4]{Section \ref{sec:4}}, 
the spin three and spin four respectively are studied in details, insisting on the gauge 
fixing procedure and identifying the Schouten tensor;
\item In \hyperref[sec:5]{Section \ref{sec:5}}, we present the spectrum of one-forms needed to describe 
the conformal spin-$s$ fields, as well as the first order equations that there are subject to;
\item Finally, \hyperref[sec:conclusions]{we conclude} the present note by commenting on 
a way towards a nonlinear completion of the unfolded system.
\end{itemize}

\section{Unfolded spin-$s$ formulation based on $\gl(5,\R)$}
\label{sec:2}

We introduce a one-form valued in the rectangular two-row representation 
$(s-1,s-1)$ of $\gl(5, \R)$, i.e. $A^{M(s-1),N(s-1)}\,$
such that $A^{M(s-1),MN(s-2)}\equiv 0\,$ and we use conventions 
whereby indices having the same name (and the same position, 
whether covariant or contravariant) are implicitly symmetrised 
with strength one. Moreover, two groups of indices separated by a comma belong to the first and second row of an irreducible Young tableau, respectively. By using commuting variables $(z_M, w_N)$ it can be packaged in the single master-field
\begin{equation}
A:=A^{M_1...M_{s-1},N_1...N_{s-1}}\,z_{M_1}...z_{M_{s-1}}\,w_{N_1}...w_{N_{s-1}},    
\label{ConnecAs}
\end{equation}
where $M=(m,+,-)$ and with light-cone directions defined as
$$
z_\pm:=z_3\pm z_{0'}\;\quad \de^\pm:=\tfrac12(\de^3\pm \de^{0'})\;,\quad \eta_{+-}=2\;, \quad \eta^{+-}=\tfrac{1}{2}\ ,
$$
and the $\so(1,2)$ index taking values $m=0,1,2\,$.
The rank and irreducibility conditions are expressed on $A$ as
\begin{equation}
(z_M\de_z^M-s+1)A=0=(w_M\de_w^M-s+1)A\;,\quad z_M\de^M_w A=0\;.    
\end{equation}
We introduce the $\so(2,3)$ generators
\begin{equation}
J_{MN}:=2\,(z_{[M}\de^z{}_{N]}+w_{[M}\de^w{}_{N]})\, ,
\end{equation}
and, by identifying the flat space translation generator as $P_m:=J_{m+}$, we define the following nilpotent operator
\begin{equation}
D_0:=\dd+h^m\,P_m=\dd+h^m\,\big[2\,z_m\de^-_z+2\,w_m\de^-_w-z_+\de_m^z-w_+\de_m^w\big] \;,   
\end{equation}
and we introduce the weight operator\footnote{The weight $\Delta$ actually corresponds to the conformal weight of the \emph{generator} associated to a given one-form, the latter receiving the opposite weight.}
\begin{equation}
\Delta:=z_+\de^+_z+w_+\de^+_w-z_-\de^-_z-w_-\de^-_w\;.    
\end{equation}
The highest weight $(s-1)$ component is identified with the vielbein 
$e^{m(s-1)}:=A^{m(s-1),+(s-1)}\,$ while the lowest weight\footnote{Notice that 
the part $\sigma^- = h^m\,P_m$ of $D_0$ raises 
the weight by one.} $-(s-1)$ component $F^{m(s-1)}:=A^{m(s-1),-(s-1)}\,$ 
is glued to the Cotton zero-form. 
We propose then the following field equations
\begin{equation}\label{spin s eq almost gl5}
\begin{split}
& D_0A^{M(s-1),N(s-1)}=0\;,\quad    \Delta > -(s-1)\;,\\
& \dd F^{m(s-1)}=h^r\wedge h^s\,\epsilon_{rsn}\,\Phi^{m(s-1)n}\;,\quad\Delta=-(s-1)\;,
\end{split}    
\end{equation}
where we take the zero form $\Phi^{m(s)}$ to be traceless, and the left hand side of the first equation above is understood as the $z_{M_1}...z_{M_{s-1}}\,w_{N_1}...w_{N_{s-1}}$ component of $D_0 A$ with weight $\Delta \neq -(s-1)$.
Having introduced a $\gl(5, \R)$-valued master field, it is natural to expect that only its traceless $\so(2,3)$ part will be sourced by the zero-form. In order to show this explicitly, let us rewrite the field equations in a fully $\gl(5)$ covariant way. 
The differential $D_0$ can be written as
\begin{equation}
D_0=\dd+\Omega=
\dd+\tfrac{1}{2}\,\Omega^M{}_N\,J^N{}_M \equiv 
\dd+\Omega^M{}_N\,\big[z_M\de_z^N+w_M\de_w^N\big]
\end{equation}
with the only non zero component being $\Omega_{m-}=-\Omega_{-m}=2\,h_m\,$. 
As is customary in geometric formulations of gravity 
and higher spin theories \cite{Preitschopf:1998ei}, one introduces a compensating vector $V_M$ that, 
in this conformal setting, is taken  
to be null and pointing in the $+$ direction 
in a preferred frame:
\begin{equation}\label{vectorV}
V_M\,V^M=0\;,\quad  V_M=\delta_{M,+} \;.   
\end{equation}
By means of the compensating vector one can define a covariantized 
background vielbein as
\begin{equation}
H_M:=D_0 V_M=\Omega_M{}^N\,V_N=(h_m,0,0)\;,    
\end{equation}
that can be used to write all the field equations at once and in manifest $\gl(5)$ covariant form as
\begin{equation}
\label{spin s eq gl5}
D_0A^{M(s-1),N(s-1)}=H_R\wedge H_S\,\Phi^{M(s-1)R,N(s-1)S}\;.  
\end{equation}
In the above equation the Cotton zero form $\Phi^{M(s),N(s)}$ is transverse, traceless and has weight $-(s-1)$
\begin{equation}
V_R\,\Phi^{RM(s-1),N(s)}=0\;,\quad \hat\Phi^{M(s), N(s-2)}=0\;,\quad \Delta(\Phi)=-(s-1)  \;, 
\end{equation}
where $\hat\Phi^{M(s), N(s-2)} := \eta_{PQ} \Phi^{M(s), PQN(s-2)}\,$.
[In general, we use a hat to denote the trace.] 
Transversality with respect to the vector \eqref{vectorV} ensures that $\Phi^{M(s),N(s)}$ 
does not have $+$ components, and the weight condition fixes the number of $-$ components 
to be $s-1\,$, such that the only non zero components are
\begin{equation}
\Yboxdim{9pt}
\Phi^{m(s),n-\ldots-}\sim \gyoung(_5s,;)\;.
\end{equation}
At this stage, by identifying the symmetric spin $s$ tensor $\Phi^{m(s)}$ as
\begin{equation}
\Phi^{m_1...m_s}= -\tfrac12\,\Phi^{m_1...m_{s-1}p,q-...-}\epsilon_{pq}{}^{m_s} \ ,
\end{equation}
one recovers the field equations in the form \eqref{spin s eq almost gl5}. 
The tensor $\Phi^{m_1...m_s}$ is symmetric and traceless by virtue of the 
tracelessness and $\mathfrak{gl}(3)$ irreducibility of $\Phi^{m(s),n}\,$. 
It is now easy to see that the trace part of the field content does not glue 
to the zero form: let us split the master field $A$ into its traceless $\so(2,3)$ 
part and its trace
\begin{equation}
\begin{split}
A^{M(s-1),N(s-1)} &= X^{M(s-1),N(s-1)}+Z^{M(s-1),N(s-3)}\eta^{NN}+(-1)^{s-1}\,Z^{N(s-1),M(s-3)}\eta^{MM}\\
&-(s-1)\,Z^{M(s-2)N,N(s-3)}\eta^{MN}\;,
\end{split}
\end{equation}
By defining a trace operation\footnote{The trace on $z$ variables and the mixed one are all related by irreducibility.} as
\begin{equation}
{\rm Tr}:=\frac{\de^2}{\de w_M\de w^M}   \ , 
\end{equation}
one can see that it commutes with $D_0$ and, by applying it on the field equations \eqref{spin s eq gl5}, one can deduce
\begin{equation}\label{general system}
\begin{split}
& D_0 X^{M(s-1),N(s-1)}=H_R\wedge H_S\,\Phi^{M(s-1)R,N(s-1)S}\;,\\
& D_0 Z^{M(s-1),N(s-3)}=0\;.
\end{split}    
\end{equation}
Hence, the one-forms $Z^{M(s-1),N(s-3)}$ give a contractible cycle.  

In the weight $\Delta=-1$ sector, by fixing to zero the pure-trace part of the $\mathfrak{gl}(5,\mathbb{R})$  connection, 
we show (see \hyperref[sec:5]{Section \ref{sec:5}}) that it is possible  
to leave out a single $\gl(3,\mathbb{R})$ component
\begin{equation}
\Yboxdim{9pt}
f^{m(s-1),n(s-2)} \sim \gyoung(_5\s,_4\su)\;.    
\end{equation}
By dimensional analysis, we know that the spin-$s$ Schouten tensor $P_{\mu(s)}$ 
is contained in the connection 
$f^{m(s-1),n(s-2)}\,$ through the dualisation 
\begin{align}
\tilde f^{r(s-2);p}:= \epsilon^{r_1 m_1 n_1}\ldots \epsilon^{r_{s-2} m_{s-2} n_{s-2}}\;  f^p{}_{m(s-2),n(s-2)}\;, 
\end{align}
where we recall that indices with the same name are implicitly 
symmetrised with strength one, and a semicolon is used to separate indices without symmetry relations.  
Indeed, in the general case the spin-$s$ dreibein has weight $(s-1)\,$, hence the Schouten tensor, 
that is expressed as $s$ derivatives of it, has to be contained in a one-form of weight $-1$.

We know that the $Z$ fields form a contractible cycle; 
therefore they can effectively be set to zero. 
In this paper we want to adopt a manifest 3D presentation of fields, 
both in the basis and in the fiber. 
Then, extracting $\so(1,2)$-valued one-forms out of 
an $\so(2,3)$-valued one form leaves us with totally symmetric and 
traceless tensors, which hides 
a hierarchy of connections like the one we have in the Lopatin-Vasiliev 
treatment of higher-dimensional systems. Therefore, in order to clearly 
organise the spectrum of connections with respect to the conformal 
weight, distinguishing them on the basis of their 3-dimensional fiber structure, 
we adopt the same strategy as followed in \cite{Boulanger:2014vya} 
and use a $\gl(3)$-valued set of one-forms that exhibit the same 
pattern as in the Lopatin-Vasiliev system, remembering that they 
can be rederived from a single $\gl(5)$-valued 2-row connection, for 
which the branching to $\gl(3)$ is direct. 
Another fundamental reason for our choice of $\gl(3)$-valued connections 
instead of $\so(1,2)$-valued ones is that we want to take advantage of 
Hodge dualisations in 3D, both in the basis and in the fiber, distinguishing 
two quantities that are related by a dualisation. 
 
In the next sections, we display the unfolded equations and the occurrence of the 
Schouten tensor in the particular cases $s=3$ and $s=4\,$. 
After the experience acquired on these two cases, we treat the general spin-$s$ case, 
focusing on the sector relevant for the spin-$s$ Schouten tensor.

\section{Spin 3 case}
\label{sec:3}
In this section we are going to focus on the spin three case, where a detailed analysis in the unfolded framework was carried out in \cite{Linander:2016brv}, while the spin four case will be treated in detail in section \ref{sec:4} and the general spin-$s$ system will be analysed in section \ref{sec:5} . The connection one-form is now valued in the Young tableau of $\gl(5, \R)\,$ with two rows of length two: $A^{MN,PQ}\,$. The system \eqref{spin s eq gl5} splits as
\begin{align}
& \dd e^{mn}-h_p\,\omega^{mn,p}+2\,h^{(m}b^{n)}=0\\
& \dd \omega^{mn,p}-2h_q\,X^{mn,pq}+2\big(h^p\,f^{mn}-h^{(m}\,f^{n)p}\big)-2h^{(m}\,b^{n),p}=0\\
& \dd X^{mn,pq}+2\big[h^{(m|}\,f^{pq,|n)}+h^{(p|}\,f^{mn,|q)}\big]=0\\
& \dd b^m+h_n\,\big[f^{mn}+b^{m,n}\big]+4\,h^m\,k=0\\
& \dd b^{m,n}-2\,h_p\,f^{p[m,n]}-6\,h^{[m}\,k^{n]}=0\\
& \dd k-h_m\,k^m=0\\
& \dd k^m-2\,h_n\,F^{mn}=0\\
& \dd f^{mn}-h_p\,f^{mn,p}-2h^{(m}\,k^{n)}=0\\
& \dd f^{mn,p}+4\,\big[h^p\,F^{mn}-h^{(m}\,F^{n)p}\big]=0\\
& \dd F^{mn}=h^r\, h^s\,\epsilon_{rsp}\,\Phi^{mnp}
\end{align}
where we defined
\begin{equation}
\begin{split}
&\Delta = 2\;:\quad e^{mn}:=A^{mn,++}\;,\\
&\Delta = 1\;:\quad \omega^{mn,p}:=2\,A^{mn,p+}\;,\quad b^m:=2\,A^{m-,++}\\
&\Delta = 0\;:\quad X^{mn,pq}:=A^{mn,pq}\;,\quad f^{mn}:=2\,A^{mn,+-}\;,\quad b^{m,n}:=4\,A^{+[m,n]-}\;,\quad k:=A^{++,--}\\
&\Delta = -1\;:\quad f^{mn,p}:=2\,A^{mn,p-}\;,\quad k^m:=2\,A^{m+,--}\\
&\Delta = -2\;:\quad F^{mn}:=A^{mn,--}\;,
\end{split}    
\end{equation}
As we have shown in the general analysis above, only the $\so(2,3)$ part of the connection is sourced by the zero form, {i.e.} the fields \begin{equation}
Z^{MN}:=A^{MN,P}{}_P=A^{MN,p}{}_p+4\,A^{MN,+-}  \end{equation}
form a decoupled subsystem and can be consistently set to zero, yielding the identifications
\begin{equation}
\begin{split}
&\hat e=0\;,\quad b^m=\tfrac12\,\hat\omega^m\;,\quad f^{mn}=-\tfrac12\,\hat X^{mn} \;;\\
&k=\tfrac14\,\hat f=-\tfrac18\,\hat{\hat{X}} \;,\quad k^m=\tfrac12\,\hat f^m\;,\quad \hat F=0\;,
\end{split}    
\end{equation}
where hats denote traces, taken as $\hat{\omega}^m:=\omega^p{}_p{}^{,m}\;$, $\hat{f}^m:=f^p{}_p{}^{,m}$ and $\hat{X}^{mn}:=X^{mn,p}{}_p\,$, the others being unambiguous.
The field equations in terms of the independent fields thus read
\begin{align}
&\label{demn} \dd e^{mn}-h_p\,\omega^{mn,p}+2\,h^{(m}b^{n)}=0\\
&\label{domegamn,p} \dd \omega^{mn,p}-2h_q\,X^{mn,pq}-\big(h^p\,\hat X^{mn}-h^{(m}\,\hat X^{n)p}\big)-2h^{(m}\,b^{n),p}=0\\
& \label{dXmn,pq} \dd X^{mn,pq}+2\big[h^{(m|}\,f^{pq,|n)}+h^{(p|}\,f^{mn,|q)}\big]=0\\
& \dd b^{m,n}-2\,h_p\,f^{p[m,n]}-3\,h^{[m}\,\hat f^{n]}=0\\
&\label{dfmn,p} \dd f^{mn,p}+4\,\big[h^p\,F^{mn}-h^{(m}\,F^{n)p}\big]=0\\
& \dd F^{mn}=h^r\, h^s\,\epsilon_{rsp}\,\Phi^{mnp}
\end{align}
with the trace constraints
$\hat e=0\;,\; \hat F=0\;,\; \hat\Phi^m=0 \;$,  and the definition $b^m:=\tfrac12\,\hat\omega^m\,$.  
The gauge transformations
\begin{align}
& \delta e^{mn}=\dd\varepsilon^{mn}-h_p\,\varepsilon^{mn,p}+2\,h^{(m}\sigma^{n)}\\
&\delta\omega^{mn,p}=\dd\varepsilon^{mn,p}-2h_q\,\varepsilon^{mn,pq}-\big(h^p\,\hat \varepsilon^{mn}-h^{(m}\,\hat \varepsilon^{n)p}\big)-2h^{(m}\,\sigma^{n),p}\\
&\delta X^{mn,pq}=\dd\varepsilon^{mn,pq}+2\big[h^{(m|}\,\lambda^{pq,|n)}+h^{(p|}\,\lambda^{mn,|q)}\big]\\
&\label{gaugetr bmn} \delta b^{m,n}=\dd\sigma^{m,n}-2\,h_p\,\lambda^{p[m,n]}-3\,h^{[m}\,\hat\lambda^{n]}\\
& \label{gaugetr fmnp}\delta f^{mn,p}=\dd\lambda^{mn,p}+4\,\big[h^p\,\Lambda^{mn}-h^{(m}\,\Lambda^{n)p}\big]\\
& \delta F^{mn}=\dd\Lambda^{mn}\;, 
\end{align}
are then subject to the same constraints:
$\hat \varepsilon=0\;,\; \hat \Lambda=0\;$, and $\sigma^m:=\tfrac12\,\hat\varepsilon^m\;$.   

\subsection{The $B$-gauge}

In order to simplify the analysis of the system, let us consider the gauge transformations of the $B$-fields:\footnote{Here and in the following a vertical bar is used to separate the foot index of a one-form.} 
\begin{equation}
\begin{split}
\delta b_{\mu\vert m} &= \de_\mu\sigma_m -\sigma_{\mu,m}-\tfrac12\,\hat\varepsilon_{\mu m}-\tfrac12\,h_{\mu m}\,\hat{\hat\varepsilon}\;,\\
\delta b_{\mu\vert m,n} &= \de_\mu\sigma_{m,n}-2\,\lambda_{\mu[m,n]}-3\,h_{\mu [m}\,\hat\lambda_{n]}
\end{split}    
\end{equation}
By expressing the fields in terms of $\gl(3,\R)$ irreducibles, {i.e.}
\begin{align}
b_{\mu|m}&=b_{(\mu| m)}+b_{[\mu|n]}\\
b_{\mu|m,n}&=\tfrac23\,\big(b_{\mu m,n}-b_{\mu n,m}\big)-\tfrac16\,\epsilon_{\mu mn}\,b\;,
\end{align}
where $b_{\mu m,n}:=b_{(\mu|m),n}$ and $b:=\epsilon^{\mu mn}\,b_{\mu|m,n}$, one can see that by using completely the gauge parameters
$\hat\varepsilon^{mn}$, $\sigma^{m,n}$ and $\lambda^{mn,p}$ one can set
\begin{equation}\label{B gauge}
b_{\mu|m}=0=b_{\mu m,n} \;,    
\end{equation}
leaving in particular the residual gauge parameters
\begin{align}
\sigma^{m,n}_{\rm res} &= \de^{[m}\sigma^{n]}\;,\\
\lambda^{mn,p}_{\rm res} &= \tfrac23\,\de^{(m}\sigma^{n),p}_{\rm res}-\tfrac29\,\big[\eta^{mn}\de\cdot\sigma^{p}_{\rm res}-\eta^{p(m}\de\cdot\sigma^{n)}_{\rm res}\big]\;,
\end{align}
where we defined $\de\cdot\sigma^{n}_{\rm res}:=\de_m\sigma^{m,n}_{\rm res}\,$. In this gauge one has $\hat\omega^m=0$ and the corresponding field equations for $b^m$ and $b^{m,n}$ reduce to the following constraints:
\begin{equation}\label{constr fmn fmn,p}
\begin{split}
& h_n\,\hat X^{mn}+\tfrac13\,h_n\,h_p\,\epsilon^{mnp}\,b+h^m\,\hat{\hat X}=0\quad\Rightarrow\quad b=0\;,\quad  h_n\,\hat X^{mn}+h^m\,\hat{\hat X}=0\;,\\
& 2\,h_p\,f^{p[m,n]}+3\,h^{[m}\,\hat f^{n]}=0\;.
\end{split}
\end{equation}
As previously recalled, the Schouten tensor has weight $-1$; therefore, in the spin three case it has to be contained in $f^{mn,p}\,$, so we turn now to decompose it into its $\gl(3,\R)$ irreducibles and to analyse its gauge symmetries. In order to find the rank three symmetric tensor representing the Schouten, it is convenient to dualise the one form $f^{mn,p}$:
\begin{equation}
\tilde f^{m;n}:=\epsilon^m{}_{rs}\,f^{nr,s}\;,\quad \eta_{mn}\,\tilde f^{m;n}\equiv0\;,\quad\Leftrightarrow\quad f^{mn,p}=-\tfrac23\,\tilde f^{r;(m}\epsilon_r{}^{n)p}\;,    
\end{equation}
where a semicolon separates groups of indices without symmetry relations. The symmetric and antisymmetric parts of $\tilde f^{m;n}$ are given by the duals of the traceless\footnote{A check on a field denotes the traceless projection on the displayed indices.} and trace parts of $f^{mn,p}$ as
\begin{equation}
\tilde f^{m;n}= S^{mn}+A^{m,n}\;,\quad S^{mn}:= \tilde f^{(m;n)}=\epsilon^m{}_{rs}\,\check f^{nr,s}\;,\quad A^{m,n}:=\tilde f^{[m;n]}=\tfrac34\,\epsilon^{mn}{}_p\,\hat f^p\;.    
\end{equation}
The gauge symmetry of the dualised fields can be read off from \eqref{gaugetr fmnp}:
\begin{equation}\label{gaugetr Smn and Amn}
\begin{split}
\delta S^{mn}&=d\tilde\lambda^{(m;n)}_{\rm res}-6\,h_p\,\tilde\Lambda^{mn,p}\;,\\
\delta A^{m,n}&=d\tilde\lambda^{[m;n]}_{\rm res}-4\,h_p\,\tilde\Lambda^{p[m,n]}\;,
\end{split}    
\end{equation}
where we have defined the dual gauge parameters as
\begin{equation}
\begin{split}
\tilde\Lambda^{mn,p} &:= \Lambda^{q(m}\,\epsilon_q{}^{n)p}\;,\quad \eta_{mn}\,\tilde\Lambda^{mn,p}\equiv0\;,\\
\tilde\lambda^{m;n}_{\rm res} &:= \epsilon^m{}_{rs}\,\lambda^{nr,s}_{\rm res}=\tfrac12\,\epsilon^m{}_{rs}\,\de^n\de^r\sigma^s-\tfrac16\,\epsilon^{mnr}\,\big(\Box\sigma_r-\de_r\de\cdot\sigma\big)\;,\quad \eta_{mn}\,\tilde\lambda^{m;n}_{\rm res}\equiv0\;.
\end{split}    
\end{equation}

\subsection{The Schouten tensor}

Treating form and fiber indices on the same footing by means of the background dreibein, the tensor $S_{\mu|\nu\lambda}$ decomposes into a totally symmetric, traceful, part $P_{\mu\nu\lambda}$, and a traceless hook $H^{(s)}_{\mu\nu,\lambda}$ as follows
\begin{align}\label{S decomposition}
S_{\mu|\nu\lambda} &= P_{\mu\nu\lambda}-\tfrac12\,(\eta_{\nu\lambda}\,\hat P_\mu-\eta_{\mu(\nu}\,\hat P_{\lambda)})+H^{(s)}_{\nu\lambda,\mu}\;,\\[3mm]
P_{\mu\nu\lambda}&:=S_{(\mu|\nu\lambda)}\;,\quad \hat P_\lambda:=\eta^{\mu\nu}\,P_{\mu\nu\lambda}=\tfrac23\,\hat S_\lambda=:\eta^{\mu\nu}\,S_{\mu|\nu\lambda}\;,\\
H^{(s)}_{\nu\lambda,\mu}&:=\tfrac23\,(S_{\mu|\nu\lambda}-S_{(\nu|\lambda)\mu})+\tfrac13\,(\eta_{\nu\lambda}\,\hat S_\mu-\eta_{\mu(\nu}\,\hat S_{\lambda)})\;.
\end{align}
Inserting the above decomposition into the gauge transformation \eqref{gaugetr Smn and Amn} one gets
\begin{align}
\delta P_{\mu\nu\lambda} &= \tfrac12\,\de_{(\mu}\de_\nu w_{\lambda)}\;,\label{gaugetr Pmnp}\\[2mm]
\delta H^{(s)}_{\nu\lambda,\mu} &= \tfrac16\,[\de_\mu\de_{(\nu}w_{\lambda)}-\de_\nu\de_\lambda w_\mu]-\tfrac{1}{12}\,[\eta_{\mu(\nu}\,\Box w_{\lambda)}-\eta_{\nu\lambda}\,\Box w_\mu]-6\,\tilde\Lambda_{\nu\lambda,\mu}\;,\label{gaugetr H(s)mn,p}
\end{align}
where $w^\mu:=\epsilon^{\mu\nu\lambda}\de_\nu\sigma_\lambda\,$. This shows that $P_{\mu\nu\lambda}$ has indeed the correct transformation property for the Schouten tensor, and from the second equation one can see that the parameter $\tilde\Lambda_{\nu\lambda,\mu}$ can be used to gauge away the hook component $H^{(s)}_{\nu\lambda,\mu}\,$, leaving a residual parameter that can be read off from \eqref{gaugetr H(s)mn,p}.\\
Turning to the antisymmetric one-form $A^{m,n}$, it can be decomposed into a traceful hook $H^{(a)}_{\mu\nu,\lambda}$ and a singlet $s$ as
\begin{align}\label{A decomposition}
A_{\mu|\nu,\lambda}=\tfrac23\,(H^{(a)}_{\mu\nu,\lambda}-H^{(a)}_{\mu\lambda,\nu})-\tfrac16\,\epsilon_{\mu\nu\lambda}\,s\;,\\[2mm]
H^{(a)}_{\mu\nu,\lambda}:=A_{(\mu|\nu),\lambda}\;,\quad s:=\epsilon^{\mu\nu\lambda}\,A_{\mu|\nu,\lambda}\;,
\end{align}
and one can see that $A^{m,n}$ does not have any shift symmetry left. At this point one can use the second constraint of \eqref{constr fmn fmn,p}, namely
$$
2\,h_p\,f^{p[m,n]}+3\,h^{[m}\hat f^{n]}=0\;,
$$
that by dualisation on $m,n$ gives
\begin{equation}
h_n\,S^{mn}+3\,h_n\,A^{m,n}=0\;.    
\end{equation}
Now, making use of the decompositions \eqref{S decomposition} and \eqref{A decomposition} one can solve
\begin{equation}
s=0\;,\quad
H^{(a)}_{\mu\nu,\lambda}=-\tfrac14\,\big[\eta_{\mu\nu}\hat P_\lambda-\eta_{\lambda(\mu}\hat P_{\nu)}\big]    
\end{equation}
in the gauge $H^{(s)}_{\mu\nu,\lambda}=0\,$. At this stage the one-form $\tilde f^{m;n}$ is entirely expressed in terms of the Schouten tensor as
\begin{equation}\label{tildefm;n(P)}
\tilde f_{\mu|m;n}=P_{\mu mn}-\tfrac12\,\eta_{mn}\,\hat P_\mu+\tfrac12\,h_{\mu n}\,\hat P_m  \;,  
\end{equation}
making it a convenient starting point to solve the field equations.

\subsection{The Cotton tensor}

Let us start by considering the gluing equation:
$$
\dd F^{mn}=h^r\, h^s\,\epsilon_{rsp}\,\Phi^{mnp}\;.
$$
By opening the form indices and dualising one finds
\begin{equation}\label{Phi(F)}
\Phi^{mnp}=-\tfrac{1}{2}\,\epsilon^{\mu\nu p}\,\de_\mu F_{\nu|}{}^{mn}\;.    
\end{equation}
The solution forces the right hand side to be totally symmetric and traceless, hence imposing some differential constraints on $F^{mn}\,$. Instead of analysing them we will continue solving the chain of equations, as these constraints will become identically satisfied.
We turn now to the next equation  \eqref{dfmn,p}, that reads 
\begin{equation}
\dd\tilde f^{m;n}-6\,\epsilon^m{}_{pq}\,h^p\,F^{qn}=0\;,    
\end{equation}
in terms of the dualised field.
It is solved by
\begin{equation}
F_{\mu|mn}=\tfrac16\,\epsilon^{pq}{}_m\,\de_p\tilde f_{q|\mu;n}-\tfrac{1}{12}\,h_{\mu m}\,\epsilon^{pqr}\,\de_p\tilde f_{q|r;n}\;,   
\end{equation}
that in terms of \eqref{tildefm;n(P)} becomes
\begin{equation}
F_{\mu|mn}=\tfrac16\,\epsilon^{pq}{}_m\,\de_pP_{q\mu n}+\tfrac16\,h_{\mu[m}\,\epsilon^{pq}{}_{n]}\,\de_p\hat P_q-\tfrac{1}{12}\,\epsilon_{mn}{}^p\,\de_p\hat P_\mu    \;.
\end{equation}
The right hand side of the above expression is manifestly traceless in $(mn)$, but the equation forces it to be also symmetric. This means that it has to be annihilated by contraction with $\epsilon^{mn}{}_\nu$, yielding the Bianchi identity\footnote{At this level this is an integrability condition, it becomes an identity once the Schouten tensor is expressed in terms of the metric-like field.}
\begin{equation}\label{SchoutenBianchi}
\de^{\rho} P_{\mu\nu\rho}-2\,\de_{(\mu}\hat P_{\nu)}=0    
\end{equation}
and allowing to write
\begin{equation}
F_{\mu|mn}=\tfrac16\,\epsilon_{pq(m}\,\de^pP^q{}_{n)\mu}   \;. 
\end{equation}
The known relation between the Cotton and Schouten tensors is then recovered by using the above expression in \eqref{Phi(F)}, together with \eqref{SchoutenBianchi}:
\begin{equation}\label{Phi(P)}
\Phi_{\mu\nu\lambda}=-\tfrac{1}{12}\,\big[\Box P_{\mu\nu\lambda}-3\,\de_{(\mu}\de_\nu\hat P_{\lambda)}\big]\;.    
\end{equation}

\subsection{The Lopatin-Vasiliev chain}

At this point one can focus on the trace\footnote{The field $X^{mn,pq}$ is pure trace, hence there is no loss of information.} of the equation \eqref{dXmn,pq} that reads 
\begin{equation}
\dd \hat X^{mn}-\tfrac43\,h_p\,\tilde f^{r;(m}\epsilon_r{}^{n)p}-\tfrac43\,h^{(m}\epsilon^{n)}{}_{rs}\,\tilde f^{r;s}=0    
\end{equation}
in terms of the Schouten one-form $\tilde f^{m;n}$ and, by using \eqref{tildefm;n(P)}, it can be solved as
\begin{equation}
P_{mnp}=\tfrac34\,\epsilon^{\mu\nu}{}_p\,\de_\mu \hat X_{\nu|mn}-\tfrac38\,\eta_{mn}\,\epsilon^{\mu\nu}{}_p\de_\mu\hat{\hat X}_\nu
+\tfrac34\,\eta_{p(m}\,\epsilon^{\mu\nu}{}_{n)}\de_\mu\hat{\hat X}_\nu\;,
\end{equation}
where the symmetry properties of the right hand side will be manifest after expressing it in terms of the fundamental dynamical field residing in $e^{mn}\,$.
By virtue of the first constraint in \eqref{constr fmn fmn,p} the only independent part of $\hat X_{\mu|mn}$ is its totally symmetric projection, yielding
\begin{equation}
\hat X_{\mu|mn}=\hat X_{\mu mn}+\tfrac23\,\big[h_{\mu(m}\,\hat{\hat X}_{n)}-\eta_{mn}\,\hat{\hat X}_{\mu}\big]\;,\quad \hat X_{\mu mn}:=\hat X_{(\mu|mn)}\;,  
\end{equation}
where the traces of $\hat X_{\mu|mn}$ are related as
$$
\hat{\hat X}_\mu =\tfrac13\,\eta^{mn}\,\hat X_{m|n\mu}=\tfrac37\,\eta^{mn}\,\hat X_{\mu mn}\;.
$$
From the equation \eqref{domegamn,p} in the $B$-gauge, 
it is sufficient to take a trace and symmetrise over the remaining indices to find\footnote{Recall that in the $B$-gauge $\omega^{mn,p}$ is traceless in the fiber.}
\begin{equation}\label{fmn(omega)}
\hat X_{mnp}=\tfrac13\,{\cal F}_{mnp}-\tfrac{1}{12}\,\eta_{(mn}\,\hat{\cal F}_{p)}\quad\Rightarrow\quad \hat X_{m|np}=\tfrac13\,{\cal F}_{mnp}-\tfrac{1}{12}\,\eta_{np}\,\hat{\cal F}_{m}    
\end{equation}
where we have defined
\begin{equation}\label{Fronsdal(omega)}
{\cal F}_{mnp}:=\de^q\omega_{(m|np),q}-\de_{(m}\omega^{q|}{}_{np),q}\;,    
\end{equation}
that is the well-known Fronsdal kinetic tensor, as we shall re-derive below, 
for the sake of completeness. This analysis can be found in 
\cite{Campoleoni:2012hp, Didenko:2014dwa}, for example.
We introduce the Fronsdal-like field
\begin{equation}\label{varphi(e)}
\varphi_{mnp}:=e_{(m|np)}\quad\Rightarrow\quad \hat\varphi_p=\tfrac23\,e^{m|}{}_{mp}\;,
\end{equation}
where we used $e_{p|m}{}^m=0\,$. By taking the equation \eqref{demn} for $e^{mn}$ in the $B$-gauge
\begin{equation}
\de_\mu e_{\nu|mn}-\de_\nu e_{\mu|mn}=\omega_{\nu|mn,\mu}-\omega_{\mu|mn,\nu}
\end{equation}
one gets
\begin{equation}\label{Fronsdal ingredient 1}
\omega_{(\mu|mn),\nu}=\de_\nu\varphi_{\mu mn}-\de_{(\mu|}e_{\nu|mn)}\;,    
\end{equation}
while taking a trace and symmetrising over the remaining indices gives
\begin{equation}\label{Fronsdal ingredient 2}
\omega_{p|mn,}{}^p=2\big[\de^p e_{(m|n)p}-\de_{(m}e^{p|}{}_{n)p}\big]\;.    
\end{equation}
By using the above relations \eqref{Fronsdal ingredient 1} 
and \eqref{Fronsdal ingredient 2} in \eqref{Fronsdal(omega)} 
one indeed gets the well-known expression for the Fronsdal tensor
\begin{equation}\label{Fronsdal(varphi)}
{\cal F}_{mnp}=\Box\varphi_{mnp}-3\,\de_{(m}\de\cdot\varphi_{np)}+3\,\de_{(m}\de_{n}\hat\varphi_{p)} \;,   
\end{equation}
and the Schouten tensor is finally expressed as
\begin{equation}
P_{mnp}=\tfrac14\,\epsilon_{\mu\nu (p}\,\de^\mu{\cal F}^\nu{}_{mn)}-\tfrac{1}{32}\,\eta_{(mn}\,\epsilon^{\mu\nu}{}_{p)}\de_\mu\hat{\cal F}_\nu \;.   
\end{equation}
%
\section{Spin 4 case}
\label{sec:4}

We turn now to analyse explicitly the case of spin four, that displays already some general features that are absent for spin three. Namely, we will focus on the identification of the Schouten tensor and it will be shown that the properties required 
in \cite{Henneaux:2015cda}, i.e. the precise form of the Bianchi identity and gauge transformations, are not independent from gauge fixing.

From the general system \eqref{general system} one can extract the field equations for spin four:
\begin{align}
& \dd e^{mnp}+h_q\,\omega^{mnp,q}+h^{(m}b^{np)}=0\;,\\ & \dd\omega^{mnp,q}+h_r\,\omega^{mnp,qr}+\tfrac14\,\big[h^q\,\hat \omega^{mnp}-h^{(m}\,\hat \omega^{np)q}\big]+h^{(m}b^{np),q}=0\;,\\
& \dd b^{mn}+h_p\,b^{mn,p}-\tfrac14\,h_p\,\hat \omega^{mnp}-\tfrac34\,h^{(m}\hat b^{n)}=0\;,\\
& \dd \omega^{mnp,qr}+h_s\,X^{mnp,qrs}+\big[h^{(q\rvert}\hat{X}^{mnp,\lvert r)}-\tfrac32\,h^{(m}\hat{X}^{np)(q,r)}\big]+h^{(m}b^{np),qr}=0\;,\\
& \dd b^{mn,p}+h_q\,b^{mn,pq}-\tfrac12\,h_q\big[\hat{X}^{qmn,p}-\hat{X}^{pq(m,n)}\big]-\tfrac32\,h^{(m}\hat{\hat{X}}^{n),p}+\tfrac13\,\big[h^p \hat b^{mn}-h^{(m}\hat b^{n)p}\big]=0\;,\\
& \dd X^{mnp,qrs}+h^{(s\rvert}f^{mnp,\lvert qr)}-h^{(m\rvert}f^{qrs,\lvert np)}=0\;,\\
&\label{Schoutenconstr} \dd b^{mn,pq}-\tfrac12\,h_r\big[f^{rmn,pq}+f^{rpq,mn}\big]-3\,\big[h^{(q\rvert}\hat f^{mn,\lvert p)}+h^{(m\rvert}\hat f^{pq,\lvert n)}\big]=0\;,\\
&\label{SchouteneqforBianchi} \dd f^{mnp,qr}+h^{(q\rvert}f^{mnp,\lvert r)}-\tfrac32\,h^{(m}f^{np)(q,r)}=0\;,\\
& \dd f^{mnp,q}+h^q F^{mnp}-h^{(m}F^{np)q}=0\;,\\
& \dd F^{mnp}=h^r\wedge h^s\,\epsilon_{rsq}\,\Phi^{mnpq}\;,
\end{align}
where the traces are defined as
\begin{equation}
\begin{split}
&\hat \omega^{mnp}:=\omega^{mnp,q}{}_q\;,\quad \hat b^{m}:=b^p{}_p{}^{,m}\;,\quad \hat{ X}^{mnp,q}:=X^{mnp,qr}{}_r\;,\quad \hat{\hat{X}}^{m,n}:=\hat{X}^p{}_p{}^{m,n}\;,\\
&\hat b^{mn}:=b^{mn,p}{}_p\;,\quad \hat f^{mn,p}:=f^{qmn,p}{}_q-f^{q(mn,p)}{}_q\;,\quad \hat\omega^{mn}:=\omega^{mnp,}{}_p
\end{split}
\end{equation}
and having set the $\gl(5)$ traces to zero yields the following identifications and constraints:
\begin{equation}\label{constrgl5}
\begin{split}
& b^{mn}:=\tfrac32\,\hat \omega^{mn}\;,\quad b^{mn,p}:=3\,\hat \omega^{mn,p}\;,\quad \eta_{mn}\,\omega^{mn[p,q]}=0\;,\\
&\hat b=0\;,\quad \eta_{mn}\,f^{mn[p,q]}=0\;,
\end{split}    
\end{equation}
where
\begin{equation}
\hat \omega^{mn,p}:=\eta_{qr}\,\big[\omega^{qmn,rp}-\omega^{q(mn,p)r}\big]\;.    
\end{equation}
The above system is invariant under the following gauge transformations:
\begin{align}
&\delta e^{mnp}= \dd\varepsilon^{mnp}+h_q\,\varepsilon^{mnp,q}+h^{(m}\sigma^{np)}\;,\\ 
&\delta \omega^{mnp,q}= \dd\varepsilon^{mnp,q}+h_r\,\varepsilon^{mnp,qr}+\tfrac14\,\big[h^q\,\hat \varepsilon^{mnp}-h^{(m}\,\hat \varepsilon^{np)q}\big]+h^{(m}\sigma^{np),q}\;,\\
&\delta b^{mn}= \dd\sigma^{mn}+h_p\,\sigma^{mn,p}-\tfrac14\,h_p\,\hat \varepsilon^{mnp}-\tfrac34\,h^{(m}\hat \sigma^{n)}\;,\\
&\delta \omega^{mnp,qr}= \dd\varepsilon^{mnp,qr}+h_s\,\varepsilon^{mnp,qrs}+\big[h^{(q\rvert}\hat{\varepsilon}^{mnp,\lvert r)}-\tfrac32\,h^{(m}\hat{\varepsilon}^{np)(q,r)}\big]+h^{(m}\sigma^{np),qr}\;,\\
& \delta b^{mn,p}=\dd\sigma^{mn,p}+h_q\,\sigma^{mn,pq}-\tfrac12\,h_q\big[\hat{\varepsilon}^{qmn,p}-\hat{\varepsilon}^{pq(m,n)}\big]-\tfrac32\,h^{(m}\hat{\hat{\varepsilon}}^{n),p}+\tfrac13\,\big[h^p \hat \sigma^{mn}-h^{(m}\hat \sigma^{n)p}\big]\;,\\
&\delta X^{mnp,qrs}= \dd\varepsilon^{mnp,qrs}+h^{(s\rvert}\lambda^{mnp,\lvert qr)}-h^{(m\rvert}\lambda^{qrs,\lvert np)}\;,\\
& \delta b^{mn,pq}=\dd\sigma^{mn,pq}-\tfrac12\,h_r\big[\lambda^{rmn,pq}+\lambda^{rpq,mn}\big]-3\,\big[h^{(q\rvert}\hat \lambda^{mn,\lvert p)}+h^{(m\rvert}\hat \lambda^{pq,\lvert n)}\big]\;,\\
& \delta f^{mnp,qr}=\dd\lambda^{mnp,qr}+h^{(q\rvert}\lambda^{mnp,\lvert r)}-\tfrac32\,h^{(m}\lambda^{np)(q,r)}\;,\\
& \delta f^{mnp,q}=\dd\lambda^{mnp,q}+h^q \Lambda^{mnp}-h^{(m}\Lambda^{np)q}\;,\\
& \delta F^{mnp}=\dd\Lambda^{mnp}\;,
\end{align}
where the gauge parameters obey the same constraints as their corresponding gauge fields as in \eqref{constrgl5}.

\subsection{The $B$-gauge}

In order to identify the Schouten tensor and simplify the system 
we start by gauging away the $B$-fields as much as possible. 
From the gauge transformation
\begin{equation}
\delta b_{\mu\rvert mn}= \de_\mu\sigma_{mn}+\sigma_{mn,\mu}-\tfrac14\,\hat \varepsilon_{mn\mu}-\tfrac34\,h_{\mu(m}\hat \sigma_{n)}   
\end{equation}
one can see that it is possible to gauge fix $b_{\mu\rvert mn}$ to zero completely\footnote{Recall that $b_{\mu\rvert mn}$ is traceless in the fiber indices and the gauge parameters obey $\hat\sigma^m=\hat{\hat{\epsilon}}^m\;$. }. In order to find the transformation law for the Schouten tensor, we are mostly interested in the residual $\sigma$-type transformations and in this case the gauge $b_{\mu\rvert mn}=0$ is preserved by 
\begin{equation}
\tilde\sigma_{m;n}^{\rm res}=\epsilon_m{}^{pq}\,\de_p\sigma_{nq}-\tfrac16\,\epsilon_{mn}{}^p\de\cdot\sigma_p\;,\quad \tilde\sigma_{m;n}:=\epsilon_m{}^{pq}\,\sigma_{np,q}\;,\quad \eta^{mn}\,\tilde\sigma_{m;n}=0\;.
\end{equation}
The gauge transformation for the next $B$-field in the dualised representation reads
\begin{equation}\label{gaugetrtildebmumn}
\delta\tilde b_{\mu\rvert m;n}=\de_\mu\tilde\sigma_{m;n}^{\rm res}-\tfrac23\,\epsilon^r{}_{n\mu}\,\tilde{\tilde{\sigma}}_{mr}-\tfrac16\,\epsilon^r{}_{mn}\,\tilde{\tilde{\sigma}}_{\mu r}-\tfrac34\,\tilde\varepsilon_{m;n\mu}+\tfrac38\,\big(\eta_{mn}\,\hat{\tilde{\varepsilon}}_\mu-3\,h_{\mu n}\,\hat{\tilde{\varepsilon}}_m\big) 
\end{equation}
where
\begin{equation}
\tilde b_{m;n}:=\epsilon_m{}^{pq}b_{np,q}\;,\quad  \tilde{\tilde{\sigma}}_{mn}:=\epsilon_m{}^{pq}\epsilon_n{}^{rs}\sigma_{pr,qs}\;,\quad \tilde\varepsilon_{m;np}:=\epsilon_m{}^{qr}\hat\varepsilon_{npq,r}   
\end{equation}
obeying
\begin{equation}
\eta^{mn}\,\tilde b_{m;n}=0=\eta^{mn}\,\tilde\varepsilon_{m;np} \;,\quad \tilde \varepsilon_{[m;n]p}=\tfrac12\,\hat{\tilde{\varepsilon}}_{[m}\eta_{n]p} \;. 
\end{equation}
The gauge field $\tilde b_{\mu\rvert m;n}$ can be decomposed into its irreducible components as
\begin{equation}
\begin{split}
\tilde b_{m;n}&=\tilde b^{(s)}_{mn}+\tilde b^{(a)}_{m,n}    \;,\\
 \tilde b^{(s)}_{\mu\rvert mn}&=\check{s}_{\mu mn}+\check h_{mn,\mu}^{(s)}-\tfrac15\,\big(\eta_{mn}v^{(s)}_\mu-3\,h_{\mu (m}v^{(s)}_{n)}\big)\;,\\
 \tilde b^{(a)}_{\mu\rvert m,n}&=\tfrac23\,\big(\check h^{(a)}_{\mu m,n}-\check h^{(a)}_{\mu n,m}\big)-\tfrac16\,\epsilon_{\mu mn}\,a+h_{\mu[m}v_{n]}^{(a)}\;,
\end{split}    
\end{equation}
where we recall that a check denotes traceless fields and the irreducibles are defined by
\begin{equation}
\begin{split}
&\check s_{\mu mn}:=\check{\tilde b}^{(s)}_{(\mu\rvert mn)}\;, \quad \check h^{(s)}_{mn,\mu}:=\tfrac23\,\big(\check{\tilde b}^{(s)}_{\mu\rvert mn}-\check{\tilde b}^{(s)}_{(m\rvert n)\mu}\big)\;,\quad v^{(s)}_n:=h^{\mu m}\,\tilde b^{(s)}_{\mu\rvert mn} \;,\\ 
&\check h^{(a)}_{\mu m,n}:=\check{\tilde b}^{(a)}_{(\mu\rvert m),n}\;,\quad v^{(a)}_n:=h^{\mu m}\,\tilde b^{(a)}_{\mu\rvert m,n}\;,\quad a:=\epsilon^{\mu mn}\tilde b^{(a)}_{\mu\rvert m,n}\;.
\end{split}    
\end{equation}
From the gauge transformation \eqref{gaugetrtildebmumn} one can derive the transformations of the various irreducible projections. In particular, one finds that the combination $v^{(s)}_m+\tfrac32\,v^{(a)}_m$ is fully gauge invariant and hence it must be zero on shell, while the remaining vector can be gauged to zero by means of $\hat{\tilde{\varepsilon}}_m\,$. Similarly one can use the traceless part of $\tilde\varepsilon_{(m;np)}$ to gauge away $\check s_{mnp}$ and the trace of $\tilde{
\tilde{\sigma}}_{mn}$ to gauge away $a$, leaving the residual parameter
\begin{equation}
\hat{\tilde{\tilde{\sigma}}}^{\rm res}=-\tfrac35\,\epsilon^{\mu mn}\,\de_\mu\tilde\sigma_{m;n}^{\rm res} \;.   
\end{equation}
The traceless hooks $\check h^{(s)}_{\mu m,n}$ and $\check h^{(a)}_{\mu m,n}$ both have a shift symmetry involving the traceless part of $\tilde{
\tilde{\sigma}}_{mn}\,$, making the gauging ambiguous; hence we choose to gauge fix to zero the linear combination 
$$
\check h^{(s)}_{\mu m,n}+\alpha\,\check h^{(a)}_{\mu m,n}\;,
$$ 
leaving 
\begin{equation}
\tilde b_{\mu\rvert m;n}=\tfrac23\,\big(\check h^{(a)}_{\mu m,n}-\check h^{(a)}_{\mu n,m}\big)-\alpha\,\check h^{(a)}_{mn,\mu} \;,   
\end{equation}
after gauge fixing.
This gauge is preserved by the residual parameter
\begin{eqnarray}\label{tildetildesigmares}
\tilde{
\tilde{\sigma}}_{mn}^{\rm res}& = &-\tfrac{4}{4-\alpha}\,\epsilon_m{}^{pq}\,
\big[\tfrac{2-\alpha}{4}\,\de_q\tilde\sigma_{n;p}^{\rm res}+\tfrac{2+\alpha}{4}\,\de_q\tilde\sigma_{p;n}^{\rm res}-\tfrac{\alpha}{2}\,\de_n\tilde\sigma_{q;p}^{\rm res}+\tfrac32\,(\alpha+1)\,\eta_{np}\de\cdot\tilde\sigma_q^{\rm res}\big] \\ && \nonumber \hspace{300pt}-\tfrac15\,\eta_{mn}\,\epsilon
^{pqr}\,\de_p\tilde\sigma_{q;r}^{\rm res}\;,
\end{eqnarray}
and this ambiguity in the gauge choice reflects the possibility of having a one-parameter family of Schouten like tensors. Among them one can find the tensor defined in \cite{Henneaux:2015cda} from its gauge transformation, for a particular choice of $\alpha$ that will be justified in the following. We turn now to the last $B$-field, whose gauge transformation in the dualised form reads
\begin{equation}
\begin{split}
&\delta \tilde{\tilde{b}}_{\mu\rvert mn}= \de_\mu\tilde{\tilde{\sigma}}_{mn}^{\rm res}-3\,\tilde{\tilde{\lambda}}_{mn;\mu}+2\,\tilde{\tilde{\lambda}}_{\mu(m;n)} \;,\\  
&\tilde{\tilde{b}}_{mn}:=\epsilon_m{}^{pq}\,\epsilon_n{}^{rs}\,b_{pr,qs}\;,\quad \tilde{\tilde{\lambda}}_{mn;p}:=\epsilon_m{}^{qr}\,\epsilon_n{}^{st}\,\lambda_{pqs,rt}\;,\quad \eta^{mp}\,\tilde{\tilde{\lambda}}_{mn;p}=0\;.
\end{split}
\end{equation}
From the above transformation one can see that the traceless symmetric and traceless hook components of $\tilde{\tilde{b}}_{\mu\rvert mn}$ can be gauged away by means of the traceless part of $\tilde{\tilde{\lambda}}_{mn;\mu}\,$. The two traces of $\tilde{\tilde{b}}_{\mu\rvert mn}$ transform as
\begin{equation}
\delta \tilde{\tilde{b}}_{\mu\rvert m}{}^m=\de_\mu\hat{\tilde{\tilde{\sigma}}}^{\rm res}-3\,\hat{\tilde{\tilde{\lambda}}}_\mu\;,\quad \delta \tilde{\tilde{b}}^m{}_{\rvert mn}=\de\cdot\tilde{\tilde{\sigma}}^{\rm res}_n+\,\hat{\tilde{\tilde{\lambda}}}_n   \;,
\end{equation}
and in particular one has
\begin{equation}
\delta \big(\tilde{\tilde{b}}_{\mu\rvert m}{}^m+3\,\tilde{\tilde{b}}^m{}_{\rvert m\mu}\big)= \de_\mu\hat{\tilde{\tilde{\sigma}}}^{\rm res}+3\,(\de\cdot\tilde{\tilde{\sigma}}^{\rm res}){}_\mu \;. 
\end{equation}
The right hand side of the above equation is generally nonzero, but it is easy to see from \eqref{tildetildesigmares} that it vanishes identically for $\alpha=-1\,$. With this choice the above linear combination of traces is gauge invariant, hence zero on shell, and the remaining trace can be gauged away by means of $\hat{\tilde{\tilde{\lambda}}}_\mu\,$, leaving as residual parameter\footnote{There is clearly a residual parameter in the hook projection as well, but it does not affect the transformation law of the Schouten tensor.} 
\begin{equation}
\tilde{\tilde{\lambda}}^{\rm res}_{(mn;\mu)}=\de_{(\mu}\tilde{\tilde{\sigma}}_{mn)}^{\rm res}\;.    
\end{equation}
This choice of $\alpha$ allows thus to set $\tilde{\tilde{b}}_{\mu\rvert mn}$ to zero completely, providing a rationale for fixing this ambiguity. Let us also notice that, with $\alpha=-1\,$, the residual parameter $\tilde{\tilde{\sigma}}^{\rm res}_{mn}$ takes the form
\begin{equation}\label{goodtildetildesigma}
\tilde{\tilde{\sigma}}^{\rm res}_{mn}=\tfrac45\,\big[\Box \sigma_{mn}-2\,\de_{(m}\de\cdot\sigma_{n)}+\tfrac56\,\eta_{mn}\,\de\cdot\de\cdot\sigma\big]\;,  
\end{equation}
that coincides with the $\nu_{mn}$ tensor of \cite{Henneaux:2015cda}.

\subsection{The Schouten tensor}

From a weight argument, that in this context just amounts to count the  number of derivatives of $e^{mnp}$, the Schouten tensor has to be contained in the one-form $f^{mnp,qr}$, whose double dual is defined as
\begin{equation}
\tilde{\tilde{f}}^{mn;p}:=\epsilon^m{}_{qr}\,\epsilon^n{}_{st}\,f^{pqs,rt}\;,\quad \tilde{\tilde{f}}^{mn;}{}_m=0  \;.   
\end{equation}
Its gauge transformation reads
\begin{equation}
\delta \tilde{\tilde{f}}_{\mu\rvert mn;p}=\de_\mu \tilde{\tilde{\lambda}}_{mn;p}^{\rm res}+\tfrac32\,\epsilon^q{}_{\mu(m}\tilde{\lambda}_{n);pq}\;,\quad \tilde{\lambda}_{m;np}:=\epsilon_m{}^{qr}\,\lambda_{npq,r}\;,
\end{equation}
and having set $\tilde{\tilde{b}}_{\mu\rvert mn}=0$ it is also subject to the constraint
\begin{equation}\label{constrtildetildef}
3\,h^p\,\tilde{\tilde{f}}_{mn;p}=2\,h^p\,\tilde{\tilde{f}}_{p(m;n)}   \end{equation}
descending from \eqref{Schoutenconstr}.
One can already see that, by defining the Schouten tensor as
\begin{equation}
P_{\mu mnp}:= \tilde{\tilde{f}}_{(\mu\rvert mn;p)}\;,  \end{equation}
it transforms as
\begin{equation}\label{Schoutentransformation}
\delta P_{mnpq}=\de_{(m}\de_{n}\tilde{\tilde{\sigma}}^{\rm res}{}_{pq)} \;,   
\end{equation}
that coincides with the transformation law given in \cite{Henneaux:2015cda} for $\tilde{\tilde{\sigma}}^{\rm res}_{mn}$ as in \eqref{goodtildetildesigma}. We turn now to show that the Schouten tensor is the only physical component of the one-form $\tilde{\tilde{f}}_{mn;p}\,$. 
First, we decompose it in the fiber indices as
\begin{equation}
\begin{split}
& \tilde{\tilde{f}}_{mn;p}=S_{mnp}+\check{H}_{mn,p}+\big(\eta_{mn}\hat S_p-\eta_{p(m}\hat S_{n)}\big)\;,\\
& S_{mnp}:=\tilde{\tilde{f}}_{(mn;p)}\;,\quad \check{H}_{mn,p}:=\tfrac23\big[\tilde{\tilde{f}}_{mn;p}-\tilde{\tilde{f}}_{p(m;n)}\big]_{\text{trace free}}\;,
\end{split}    
\end{equation}
where the fiber trace of $\tilde{\tilde{f}}_{mn;p}$ has been kept entirely in $S_{mnp}\,$. The totally symmetric and the traceless hook one-forms can be further decomposed once the form index is made explicit:
\begin{equation}\label{tildetildefdecomp}
\begin{split}
 S_{\mu\rvert mnp}&=P_{\mu mnp}+H^{(s)}_{mnp,\mu}\;,\quad {\rm where}\\
 P_{\mu mnp}&=S_{(\mu\rvert mnp)}\;,\quad H^{(s)}_{mnp,\mu}:=\tfrac34\,\big[S_{\mu\rvert mnp}-S_{(m\rvert np)\mu}\big] \;,\\[3mm]
 \check{H}_{\mu\rvert mn,p}&=\tfrac38\,\big(H^{(h)}_{mnp,\mu}+3\,H^{(h)}_{\mu mn,p}\big)+\tfrac32\,\big[\eta_{mn}\,\hat H^{(h)}_{\mu p}+h_{\mu p}\,\hat H^{(h)}_{mn}-\eta_{p(m}\,\hat H^{(h)}_{n)\mu }-h_{\mu(m}\,\hat H^{(h)}_{n)p}\big]\\
 &+\tfrac34\,\big[\eta_{mn}\,\hat H^{(h)}_{p,\mu}+h_{\mu(m}\,\hat H^{(h)}_{n),p}-\eta_{p(m}\,\hat H^{(h)}_{n),\mu }\big]\;,\quad {\rm where}\\
 H^{(h)}_{\mu mn,p}&:=\check{H}_{(\mu\rvert mn),p}\;,\quad \hat H^{(h)}_{mn}:=H^{(h)}_{p mn,}{}^p\;,\quad \hat H^{(h)}_{m,n}:=H^{(h)}_p{}^p{}_{[m,n]}\;.
\end{split}  
\end{equation}
From the constraints \eqref{constrgl5} one can see that the gauge parameter $\lambda^{mnp,q}$ has vanishing antisymmetric trace; this translates to its dual $\tilde{\lambda}^{m;np}$ being completely traceless. Using the Stueckelberg symmetry it is possible to gauge away only one traceless hook, for instance $\check{H}^{(s)}_{mnp,\mu}$, and one symmetric trace from the hooks, e.g. $H^{(s)}_{mnp,}{}^p\;$, from the decomposition \eqref{tildetildefdecomp}. Using it in the constraint \eqref{constrtildetildef} one finds
$$
\hat H^{(h)}_{mn}=\hat P_{mn}-\tfrac13\,\eta_{mn}\,\hat{\hat{P}}\;,
$$
and the other irreducible components, except for the Schouten, are all set to zero, yielding
\begin{equation}\label{gfixedtildetildef}
\tilde{\tilde{f}}_{\mu\rvert mn;p}=P_{\mu mnp}+2\,\big[\eta_{mn}\,\hat  P_{p\mu}-\eta_{p(m}\,\hat  P_{n)\mu}+h_{\mu p}\,\hat  P_{mn}-h_{\mu(m}\,\hat  P_{n)p}\big]-\big(\eta_{mn}\,h_{\mu p}-\eta_{p(m}\,h_{n)\mu }\big)\,\hat{\hat{P}}\;.    
\end{equation}
From the identity $\de_\mu\hat{\tilde{\tilde{\sigma}}}^{\rm res}+3\,\de\cdot\tilde{\tilde{\sigma}}^{\rm res}_\mu\equiv0$ and the transformation law \eqref{Schoutentransformation} it is immediate to see that
\begin{equation}
\delta \big(\de\cdot P_{mnp}-3\,\de_{(m}\hat P_{np)}\big)=0\;.    
\end{equation}
Since the quantity in brackets is fully gauge invariant, it has to vanish, providing thus the Bianchi identity for the Schouten tensor. This can be derived directly from its field equation \eqref{SchouteneqforBianchi} rewritten in double dualised form:
\begin{equation}
d\tilde{\tilde{f}}_{mn;p}+\tfrac32\,h^r\,\epsilon^q{}_{r(m}\,\tilde{f}_{n)pq}=0\;.    
\end{equation}
By opening the form indices and contracting with a further $\epsilon$ symbol one can solve
\begin{equation}
\tfrac32\,\tilde{f}_{(m\rvert n);pq}=\epsilon^{\mu\nu}{}_q\,\de_\mu\tilde{\tilde{f}}_{\nu\rvert mn;p}-\tfrac23\,\eta_{q(m\rvert}\,\epsilon^{rst}\de_r\tilde{\tilde{f}}_{s\rvert t\rvert n);p}\;,    
\end{equation}
and since the left hand side is manifestly symmetric in $(pq)$, the right hand side must be annihilated by contraction with $\epsilon^{pq}{}_u\,$ yielding
\begin{equation}
\de^q P_{mnpq}-3\,\de_{(m}\hat P_{np)}=0    
\end{equation}
upon using the decomposition \eqref{gfixedtildetildef}.
This is precisely the Bianchi identity for the spin-4 Schouten tensor 
obtained in \cite{Henneaux:2015cda}.  

\section{General, spin-$s$ case}
\label{sec:5}

In this section, we provide the unfolding of a free conformal spin-$s$ field
in terms of $\gl(3,\R)$-valued connections with appropriate trace constraints.
We then show how to produce another and direct  proof for the cohomological 
problem set up and solved in \cite{Henneaux:2015cda}. 

\subsection{Minimal spectrum}

After the expertise acquired in the previous two sections, we can now 
present the minimal spectrum of fields, i.e. with the same $\so(3)$ content as 
in \cite{Pope:1989vj}, inside a set of $\gl(3,\R)$-valued 
connections subject to a single trace constraint.
As is customary in Conformal Field Theory, we organise the spectrum 
with respect to the conformal weight $\Delta\,$. Explicitly, we find: 
\begin{align}
\Delta &= s-1 :\; \quad e^{a(s-1)} \quad {\rm{s.t.}} \quad \eta_{ab}\,e^{abc(s-3)}\equiv 0 \;;\quad s\geqslant 3 \;,
\nonumber\\
&\quad \vdots 
\nonumber\\
\Delta &= s-1-k :\; \quad \omega^{m(s-1),n(k)} \quad {\rm{s.t.}}\quad 
\underbrace{\epsilon_{pqr}\ldots \epsilon_{pqr}}_{k \;{\rm{factors}}}\;
\omega^{p(k)m(s-k-1),q(k)} \;\eta_{mm} \equiv 0 \;, \quad k\in \{1,\ldots,s-2\}\; 
\nonumber\\
&\quad \vdots 
\nonumber\\
\Delta &= 0 :\; \quad \{ X^{m(s-1),n(s-1)}, b^{m(s-2),n(s-2)}\}\; , 
\nonumber\\
& 
\nonumber\\
\Delta & = -(s-k-1):\;\; f^{m(s-1),n(k)} \quad {\rm{s.t.}} \quad
\underbrace{\epsilon_{pqr}\ldots \epsilon_{pqr}}_{k \; {\rm{factors}}} 
\;f^{p(k)m(s-k-1),q(k)} \;\eta_{mm} \equiv 0 \;, \quad k\in \{s-2,\ldots,1\}\; ,
\nonumber \\
&\quad \vdots 
\nonumber
\\
\Delta & = -s+1:\;\; F^{m(s-1)} \quad {\rm{s.t.}} \quad
\eta_{pq}\,F^{pqm(s-3)} \equiv 0 \;;\quad s\geqslant 3 \ .
\end{align}
\vspace{.5cm}

\noindent We introduce the dual connections
\begin{align}
\tilde{\omega}_{r(k);}{}^{m(s-k-1)} &:= 
\underbrace{\epsilon_{pqr}\ldots \epsilon_{pqr}}_{k \; {\rm{factors}}} 
\;\omega^{p(k)m(s-k-1),q(k)}\ ,    \quad  k\in \{1,\ldots,s-2\}\;
\\
\tilde{f}_{r(k);}{}^{m(s-k-1)} &:= 
\underbrace{\epsilon_{pqr}\ldots \epsilon_{pqr}}_{k \; {\rm{factors}}} 
\;f^{p(k)m(s-k-1),q(k)}    \ , \quad  k\in \{s-2,\ldots,1\}\;, 
\end{align}
that are constrained by 
\begin{equation}
    \eta_{pq}\,\tilde{\omega}_{r(k);}{}^{pqm(s-k-3)} \equiv 0 \equiv 
    \eta_{pq}\,\tilde{f}_{r(k);}{}^{pqm(s-k-3)}\;,\quad 
    k\in \{1,\dots ,s-3\} \ ,
\end{equation}
and we recall that groups of indices separated by a semicolon do not have symmetry relations between them.
The inverse relation between the dual fields and the original ones is 
\begin{equation}
    \omega^{m(s-1),n(k)} = (-1)^k \, \tfrac{s-k}{k}\; \tilde{\omega}^{r(k);m(s-1-k)}
    \;
    \epsilon_{r_1}{}^{mn}\ldots\epsilon_{r_k}{}^{mn}\ ,
\end{equation}
and similarly for the negative-weight sector. 
\vspace{.3cm}

The Schouten tensor is contained in the connection $f^{m(s-1),n(s-2)}\,$, 
or in terms of its dual, $\tilde{f}^{m(s-2);n}\,$, that is unconstrained. 
At level $\Delta=0\,$, we have the two unconstrained, $\gl(3,\R)$-valued 
one-form connections $\{ X^{m(s-1),n(s-1)}, b^{m(s-2),n(s-2)}\}\,$. 
By virtue of the $\gl(3)$-irreducibility
\begin{equation}
\omega^{m(s-1),m n(k-1)} \equiv 0 \equiv f^{m(s-1),m n(k-1)}\ , 
\end{equation}
the dual fields obey the following ``mixed-trace'' identity
\begin{equation}
\eta_{pq}\;\tilde{\omega}^{r(k-1)p;qm(s-k-2)}\equiv 0 \equiv 
\eta_{pq}\;\tilde{f}^{r(k-1)p;qm(s-k-2)}\ ,\quad 
k\in \{1,\ldots,s-2 \} \ .
\end{equation}
Therefore, decomposing the one-form $\tilde{\omega}^{r(k);m(s-k-1)}$ 
at weight $\Delta=s-1-k$
in terms of $\so(3)$-irreps produces the following set of one-form 
\begin{equation}
\tilde{\omega}^{r(k);m(s-k-1)}
\quad 
\underset{\so(3)}{\leadsto}
\quad 
\{ \omega_{(k)}^{a(s-1)},\omega_{(k)}^{a(s-2)},\ldots,\omega_{(k)}^{a(s-k-1)}\} \ ,
\quad k\in \{ 1,\ldots s-2\}\ .
\nonumber 
\end{equation}
The same decomposition holds for all strictly 
negative values of the conformal weight.  
At weight $\Delta=0\,$, the two connections $\{ X^{m(s-1),n(s-1)},b^{m(s-2),n(s-2)}\}$ 
decompose into the set $\so(3)$-valued one-forms
\begin{equation}
\{ X^{m(s-1),n(s-1)},b^{m(s-2),n(s-2)}\}
\quad 
\underset{\so(3)}{\leadsto}
\quad 
\{ \omega_{(s-1)}^{a(s-1)},\omega_{(s-1)}^{a(s-2)},\ldots,\omega_{(s-1)}^{a},\omega_{(s-1)}\}
\ .
\nonumber 
\end{equation}
Together, this builds the spectrum given in \cite{Pope:1989vj}. 

\subsection{Unfolded equations}
\label{subsec:Unfolded}

The above connection one-forms are subject to differential 
constraints that we present in decreasing order with respect to the conformal 
weight $\Delta_k = s-k-1\,$, $ k \in \{ 0,1,\ldots 2s-2\}\,$. 
We first notice that,
out of the three independent projections of the trace of $\omega^{m(s-1),n(k+1)}\,$, 
only two of them will contribute to make the overall $\gl(3)$ symmetry $\footnotesize \gyoung(_5\s,_3\ka)\,$.
Explicitly, 
\begin{align}
\hat\omega_{1}^{m(s-1),n(k-1)}&:=\omega^{m(s-1),n(k-1)p}{}_p \quad   {\rm and} 
\nonumber \\
\hat\omega_{2}^{m(s-2),n(k)}&:= (2k-s)\,
\omega^{m(s-2)p,}{}^{n(k)}{}_p + k(s-2)\; \omega^{m(s-3)np,}{}^{mn(k-1)}{}_p
\ .    
\end{align}
Then, the differential equations for the 1-form connections are :
\begin{itemize}
\item[(i)] 
For $k\in \{0, \ldots, s-3\}\,$,  
\begin{align}
     \dd\omega^{m(s-1),n(k)} &+ h_p\,\omega^{m(s-1),n(k)p} + 
    c_k\;h^m\,\hat\omega_{2}^{m(s-2),n(k)} 
    \nonumber \\
    & + d_k\;\left[ h^n\,\hat\omega_{1}^{m(s-1),n(k-1)} 
    - \tfrac{s-1}{s-k}\, h^m\,\hat\omega_{1}^{m(s-2)n,n(k-1)} \right] = 0 \; , 
\end{align}
where $e^{m(s-1)}\equiv \omega^{m(s-1),n(k)}$ at $k=0\,$ and 
Cartan-Frobenius integrability together with 
consistency of the first equation (at $k=0$) 
with the trace constraints on $e^{a(s-1)}$ and $\omega^{m(s-1),n}$
uniquely fixes  
\begin{equation}
c_k = -\frac{s-1}{(s-k-2)(s-k)}\;\ ,\qquad
d_k = \frac{k}{2(s-k-1)} \ .   
\end{equation}
For $k=s-2\,$, we have
\begin{align}
     \dd\omega^{m(s-1),n(s-2)} & + h_p\,X^{m(s-1),n(s-2)p} + 
    h^m\,b^{m(s-2),n(s-2)} 
    \nonumber \\
    & + \tfrac{s-2}{2}\;\left[ h^n\,\hat X^{m(s-1),n(s-3)} 
    - \tfrac{s-1}{2}\, h^m\,\hat X^{m(s-2)n,n(s-3)} \right] = 0 \; ;
\end{align}
\item[(ii)] the weight-zero sector ($k=s-1$)
\begin{equation}
\begin{split}
    & \dd X^{m(s-1),n(s-1)} + h^n\,f^{m(s-1),n(s-2)} + (-1)^{s-1}\, h^m\,f^{n(s-1),m(s-2)}  = 0 \ ,
\\[3mm]
&    \dd b^{m(s-2),n(s-2)} -\tfrac{1}{2}\, h_p\,\left[ f^{pm(s-2),n(s-2)} 
    + (-1)^{s} \,f^{pn(s-2),m(s-2)}\right] 
    \\
    & \qquad\qquad + \tfrac{s-2}{2}\,
    \left[h^n\,\hat f_2^{m(s-2),n(s-3)} + (-1)^s\,h^m\,\hat f_2^{n(s-2),m(s-3)}\right]  = 0 \ ;
\end{split}    
\end{equation}
\item[(iii)] the negative-weight sector from $\Delta = -1$ until $\Delta = -s+2\,$,
i.e. from $k=s$ until $k=2s-3\,$, 
\begin{equation}
\begin{split}
    &  \dd f^{m(s-1),n(2s-2-k)} + h^n\, f^{m(s-1),n(2s-3-k)} - \tfrac{s-1}{k-s+2}\,
    h^m\,f^{m(s-2)n,n(2s-3-k)} = 0\ ,
 \\
&    f^{m(s-1),n(0)} := F^{m(s-1)}\ ;
\end{split}
\end{equation}
\item[(iv)] and finally, for weight $\Delta = -s+1$ (i.e. $k=2s-2$): 
\begin{equation}
    \dd F^{a(s-1)} = h^p \wedge h^q \,\epsilon_{pqr}\, \Phi^{a(s-1)r}\ .
\end{equation}
\end{itemize}

\subsection{The Schouten tensor}

At this stage, we are ready to discuss the issue of the identification of the Schouten tensor, within the spectrum of one-forms, following the steps that we have already displayed for the particular cases of spin three and four. As it has been shown explicitly in the case of spin four, the occurrence at each weight of multiple fields belonging to the same irreducible $\mathfrak{so}(1,2)$ Young diagrams, makes the gauge fixing of the Stueckelberg symmetries ambiguous. For this reason, the identification of the Schouten tensor is not unique and depends on the choice of gauge fixings made in the previous steps. This is not surprising, since the Schouten tensor does not belong to the sigma minus cohomology, and therefore its identification is not an invariant concept in the unfolded framework. Hence, our focus in this section will be to find the gauge choice that gives the simplest realisation of the Schouten tensor in terms of our basis of one-form connections. 
\\
The only independent physical field among all the one-forms is the Fradkin-Tseytlin field $\phi_{\mu(s)}\,$, given by the traceless part of $\varphi_{\mu(s)}:=e_{\mu\rvert\mu(s-1)}\,$; however, in order to maintain a Weyl-covariant formulation, we choose to keep the dilation-like symmetry active and work in terms of the Fronsdal-like field $\varphi_{\mu(s)}\,$, that contains a single trace and transforms as
\begin{equation}
\delta \varphi_{\mu(s)}=\de_\mu\xi_{\mu(s-1)}+\eta_{\mu\mu}\,\sigma_{\mu(s-2)}
\end{equation}
with respect to generalised diffeomorphisms and Weyl dilations. The Schouten tensor $P_{\mu(s)}$ has to be a totally symmetric rank $s$ tensor built out of $s$ derivatives of $\varphi_{\mu(s)}\,$, hence it must be contained in the one-form $\tilde f^{m(s-2);n}$ at weight $\Delta=-1\,$. The second defining property of the Schouten tensor\footnote{See appendix C in \cite{YihaoYinPhD}.} is the invariance under $\xi$ diffeomorphisms and its transformation law under $\sigma$ dilations:
\begin{equation}
\delta P_{\mu(s)}=\de_\mu\de_\mu\nu_{\mu(s-2)}(\sigma)\;,
\end{equation}
where the tensor $\nu_{\mu(s-2)}$ is built from $s-2$ derivatives of the dilation parameter $\sigma_{\mu(s-2)}\,$.  
The final property that defines the spin-$s$ Schouten is its Bianchi identity \cite{Henneaux:2015cda}, \emph{i.e.}
\begin{equation}
\de\cdot P_{\mu(s-1)}-(s-1)\,\de_\mu\hat P_{\mu(s-2)}\equiv0\;,
\end{equation}
that, in the unfolded formulation, is equivalent\footnote{It will be justified below.} to the composite parameter $\nu_{\mu(s-2)}$ obeying
\begin{equation}
3\,\de\cdot \nu_{\mu(s-3)}+(s-3)\,\de_\mu\hat \nu_{\mu(s-4)}\equiv0\;.
\end{equation}
To begin with, let us notice that the Stueckelberg transformation of the generalised vielbein $e_{\mu\rvert m(s-1)}$ involves the traceless parameter $\tilde\varepsilon_{m;n(s-2)}\,$, whose irreducible $\mathfrak{so}(1,2)$ components consist of a rank-$(s-1)$ traceless symmetric tensor, that is a generalised Lorentz parameter used to gauge away the component of $e_{\mu\rvert m(s-1)}$ orthogonal to $\varphi_{\mu(s)}\,$, and a rank-$(s-2)$ traceless symmetric tensor that shifts the trace of $\varphi$, and thus has to be identified with $\sigma_{\mu(s-2)}\,$. This implies that the gauge fixing of the shift symmetries in the positive weight sector of the spectrum has to be done in a way such that the residual parameter $\tilde\lambda_{\rm res}^{m(s-2);n}$ at weight $\Delta=-1$ depends only on the parameter $\sigma\,$ and not on $\xi\,$. 
In general, from the cohomological analysis of the unfolded system at $\Delta=-1\,$, it is only possible to determine how many $\mathfrak{so}(1,2)$ traceless tensors survive, expressed as $s$ derivatives of $\varphi_{\mu(s)}\,$, from the one form $\tilde f_{\mu\rvert m(s-2);n}\,$. However, this does not determine which linear combination of traceless tensors of rank $s,(s-2),(s-4),...$ realises the Schouten tensor. The gauge choice that we found in this paper, called the $B$-gauge, allows to take full advantage of the basis of traceful $\mathfrak{gl}(3)$ one-forms and in particular to identify the Schouten tensor as 
\begin{equation}\label{Schoutendef}
P_{\mu(s)}:=\tilde f_{\mu\rvert\mu(s-2);\mu}\;.
\end{equation}
With this reorganisation of the off-shell spectrum of \cite{Pope:1989vj} in terms of $\mathfrak{gl}(3)$ valued one-forms we find that the Schouten tensor \eqref{Schoutendef} is the natural spin $s$ generalisation of the one identified in \cite{Horne:1988jf} for the geometrical formulation of 3D conformal gravity.
\\
For this purpose, it is convenient to write 
the negative-weight equations in the dual picture. It yields
\begin{align}
    \dd\tilde{X}^{m(s-1)} -2 h^p\,\epsilon^{m}{}_{pq}\,\tilde{f}^{m(s-2);q} &= 0 \ ,
    \\
    \dd\tilde{b}^{m(s-2)} -(s-1)\, h_p\,\tilde{f}^{m(s-2);p}
    +(s-2)\, h_p\,\tilde{f}^{pm(s-3);m}  &= 0 \ ,    
\end{align}
in the weight-zero sector ($k=s-1$), and in
the negative-weight sector from $\Delta = -1$ until $\Delta = -s+2\,$,
i.e. from $k=s$ until $k=2s-3\,$, it produces
\begin{equation}
\begin{split}
    \dd\tilde{f}^{m(2s-k-2);n(k+1-s)} 
    - \tfrac{k-s+3}{k-s+2}\,\epsilon^m{}_{pq}\, h^p\,\tilde{f}^{m(2s-k-3);n(k+1-s)q} &= 0    
\ , \quad k\in\{s, \ldots, 2s-3\} \ ,
\\
\tilde{f}^{m(0);m(s-1)} := F^{m(s-1)}\ .
\end{split}
\end{equation}
The field equations for $\tilde{b}^{m(s-2)}$ and $\tilde f^{m(s-2);n}$ (at level $k=s$) 
are invariant under the following gauge transformations: 
\begin{equation}
\begin{split}
    \delta \tilde{b}_{\mu\vert m(s-2)} & = \partial_{\mu}\tilde{\sigma}_{m(s-2)} 
    - (s-1)\,\tilde\lambda_{m(s-2);\mu} + (s-2) \,\tilde\lambda_{\mu m(s-3);m}\ ,
    \\
    \tilde{f}_{\mu\vert m(s-2);n} & = \partial_{\mu}\tilde{\lambda}_{m(s-2);n}-\tfrac{3}{2}\,
    \epsilon_{\mu}{}^q{}_m\,\tilde{\lambda}_{m(s-3);nq} \ .
\end{split}
\end{equation}
Due to the symmetries of $\tilde{\lambda}\,$, in particular that the trace 
$\tilde{\lambda}_{m(s-3)p;}{}^p \equiv 0\,$, we see that $\tilde{\lambda}\,$
possesses the same components as $\tilde{b}\,$, except for a totally symmetric, 
rank-$(s-3)$ component. Therefore, we could expect to be able to gauge away from 
$\tilde{b}$ all its components except for a linear combination of its two linearly independent 
traces $\hat{\tilde{b}}_{\mu|m(s-4)}$ and ${\tilde{b}}'_{m(s-3)}\,$, where 
the latter comes from the trace including the base index. 

In particular, setting to zero the totally symmetric component of $\tilde{b}_{\mu\vert m(s-2)}$
can be achieved with $\tilde{\lambda}_{m(s-2)\vert m}\,$, leaving a residual gauge 
symmetry with 
\begin{equation}
    \tilde\lambda^{\rm res.}_{m(s-2);m} = \partial_{m}\tilde{\sigma}_{m(s-2)}\ .
\end{equation}
Then, defining $P_{\mu(s)}:= \tilde{f}_{\mu\vert \mu(s-2);\mu}\,$, 
the gauge transformation of the totally symmetric component of 
$\tilde{f}_{\mu\vert m(s-2);n}$ gives 
\begin{equation}
    \delta P_{\mu(s)} 
    = \partial_\mu \partial_\mu\tilde\sigma_{\mu(s-2)} \ , 
\label{gaugevariaSchoutenS}
\end{equation}
which has the same form as the transformation law of the Schouten tensor.

At this stage, we argue that there is a distinguished gauge, called the $B$-gauge, 
that allows to fully gauge $\tilde{b}$ to zero by the choice of a gauge 
that identifies the two, a priori independent, traces of $\tilde b\,$. 
From the transformation laws of these two components, viz.
\begin{equation}
    \begin{split}
        \delta \hat{\tilde{b}}_{m|m(s-4)} & = \partial_{m} \hat{\tilde{\sigma}}_{m(s-4)} 
        - 3\hat{\tilde{\lambda}}_{m(s-4);m}\, ,
        \\
         \delta {\tilde{b}}'_{m(s-3)} & = (\partial \cdot \tilde \sigma)_{m(s-3)} 
         + (s-3) \hat{\tilde{\lambda}}_{m(s-4);m}\, ,
    \end{split}
\end{equation}
we see that one has the transformation  
\begin{equation}
    \delta [(s-3)\hat{\tilde{b}}_{m|m(s-4)} + 3\, {\tilde{b}}'_{m(s-3)}]
    = (s-3)\,\partial_{m}\hat{\tilde{\sigma}}_{m(s-4)} 
    + 3\, (\partial \cdot \tilde\sigma)_{m(s-3)}\ . 
\end{equation}
Therefore, if there is a residual gauge where 
$\tilde{\sigma}^{\rm res.}$ can be shown to satisfy the equation
\begin{equation}
    (s-3)\,\partial_{m}\hat{\tilde{\sigma}}^{\rm res.}{}_{m(s-4)} 
    + 3\, (\partial \cdot \tilde\sigma^{\rm res.}){}_{m(s-3)} = 0\ , 
    \label{Marcvaria}
\end{equation}
then there is a gauge-invariant combination at weight $\Delta = 0$, namely 
$(s-3)\hat{\tilde{b}}_{m|m(s-4)} + 3\, {\tilde{b}}'_{m(s-3)}\,$. 
However, there is no such gauge invariant quantity before 
weight $1-s\,$, hence it must vanish, 
thereby linking the two traces of $\tilde{b}\,$. 
In this gauge, the $B$-gauge, all of $\tilde{b}$ has been set to zero. 
The equation \eqref{Marcvaria} was shown to hold true 
in \cite{Henneaux:2015cda}. In the previous sections, 
we reproduced this equation from our framework in the cases $s=3$ and $s=4\,$.

When the residual parameter $\tilde{\sigma}$ obeys \eqref{Marcvaria}, 
it is easy to see that the combination 
$$B_{m(s-1)}:=\partial^n P_{nm(s-1)} - (s-1)\, \partial_{m}\hat{P}_{m(s-2)}$$ 
is gauge invariant. 
But again, since there is no nontrivial gauge-invariant quantity 
before the zero-form $\Phi^{a(s)}\,$, we deduce that the following 
identity is true
\begin{equation}
    \partial^n P_{nm(s-1)} - (s-1)\, \partial_{m}\hat{P}_{m(s-2)} \equiv 0 \ , 
\label{BianchiSchouten}
\end{equation}
which is the Bianchi identity for the Schouten tensor, shown in 
\cite{Henneaux:2015cda} to play a central role in the 3D off-shell 
conformal tensor calculus.
We know that the zero-from $\Phi$ is built out of $s-1$ derivatives of
the one-form $f^{m(s-2);n}\,$ and is the first gauge-invariant quantity
to possess this property, by very construction of the Cartan-integrable
system given in the previous subsection.

It is direct to see that the Bianchi identity \eqref{BianchiSchouten} 
is the necessary and sufficient condition for the complete symmetry 
of the rank-$s$ tensor
\begin{equation}
({\cal D}^{s-1}P)^{m(s-1);n}:= 
(\epsilon^{p_1q_1m_1}\,\partial_{p_1})\ldots 
(\epsilon^{p_{s-1}q_{s-1}m_{s-1}}\,\partial_{p_{s-1}})\,P_{q(s-1)}{}^n \ .
\label{SchoutenToCotton}
\end{equation}
Indeed, it is not difficult to see that this tensor is not only symmetric on 
its first $s-1$ indices but actually totally symmetric in 
its $s$ indices if and only if \eqref{BianchiSchouten} holds true, 
as was duly emphasised in \cite{Henneaux:2015cda}. 
By construction, the tensor $({\cal D}^{s-1}P)^{m(s-1);n}$ is fully 
gauge invariant, due to \eqref{gaugevariaSchoutenS}. 
Finally, the tracelessness of the above tensor is trivially true 
if one takes the trace with $\eta_{mn}\,$.
Having proven the complete symmetry of the tensor 
$({\cal D}^{s-1}P)^{m(s-1);n}$, one has the complete tracelessness.
Since we have built one invariant tensor, traceless and 
totally symmetric, out of $s-1$ derivatives of the tensor $P_{m(s)}\,$, 
we know that this can only be proportional to the zero-form $\Phi_{m(s)}\,$
that is the Cotton tensor: 
$({\cal D}^{s-1}P)^{m(s-1);n}\propto \Phi^{m(s-1)n}\,$. 
In other words, the components of the one-form connection 
$\tilde{f}_{m(s-2);n}$ that are linearly independent from 
$P_{m(s)}:=\tilde{f}_{\mu|m(s-2);m}\,h^\mu_m\,$ 
are not glued to $\Phi^{a(s-1)}$ and can therefore be set to zero.
Retrospectively, we can use the results of \cite{Henneaux:2015cda} to 
infer that the $B$ gauge can indeed be reached in the general spin-$s$ case --- 
something we have proven in detail for $s=3$ and $s=4$. 
Nevertheless, it would be interesting --- only for the technicalities involved --- 
to explicitly show that one can indeed reach the gauge where \eqref{Marcvaria} 
holds true, or equivalently where \eqref{BianchiSchouten} is true.
We hope to return to this in the near future. 

\subsection{Bridging the gap}

We can now introduce a connection that contains both 
the background one-form $\Omega$ and the spin-$s$ connection one-form $A\,$ 
both introduced in Section \ref{sec:2}: 
\begin{equation}
 \boldsymbol{W} := \Omega + A \, ,
\end{equation}
and go in the  $B$-gauge exhibited previously.  
In terms of the connection $\boldsymbol{W}\,$, 
the unfolded system describing conformal spin-$s$ tensor field reads 
\begin{equation}
    \boldsymbol{F}_{[2]}:= 
    \dd \boldsymbol{W} + \boldsymbol{W}\wedge \boldsymbol{W} = \Phi_{[2]}:=
H_R\wedge H_S\,\Phi^{M(s-1)R,N(s-1)S} z_{M_1}\ldots z_{M_{s-1}} \;w_{M_1}\ldots w_{M_{s-1}}\ .
\end{equation}
Since the connection $A$ takes its values in an abelian ideal ${\cal I}\,$,
see \eqref{ConnecAs}, one has $A^2=0$ and hence  
\begin{equation}
\boldsymbol{F}_{[2]}=\dd \Omega + \Omega^2 + D_0 A  = D_0 A \ ,    
\end{equation}
by virtue of the flatness of the background connection $\Omega\,$. 

If the Cotton tensor vanishes, $\Phi_{[2]}=0\,$, the connection $\boldsymbol{W}$ is flat: $\boldsymbol{F}_{[2]}=0\,$. 
Therefore, a zero-form gauge function $g$ can be found such that 
$\boldsymbol{W} = g^{-1} \dd g \,$. 
Decomposing $g=g_0\,\tilde{g}$ where $\Omega = g_0^{-1}\,\dd g_0\,$ 
and $\tilde{g}= 1 + \sigma$ for an infinitesimal parameter 
$\sigma\in {\cal I}$ gives 
$\boldsymbol{W} = \Omega + D_0 \sigma\,$, hence  
$A = D_0 \sigma\,$. At weight $\Delta = -1\,$, this implies that 
the Schouten tensor is pure gauge, or explicitly, 
\begin{equation}\label{puregauge}
    P_{\mu(s)} = \frac{s(s-1)}{2}\,{\partial}_{\mu}{\partial}_{\mu} \,\sigma{}_{\mu(s-2)} \ .
\end{equation}
Conversely, if the connection $\boldsymbol{W}$ is pure gauge, 
which implies \eqref{puregauge} for the Schouten tensor, one has a
vanishing Cotton tensor. Therefore, the formulation of linearised conformal 
spin-$s$ field in 3D that we have provided, in the gauge where the Schouten 
tensor is unambiguously defined, provides a direct solution for the cohomological   
problem solved in \cite{Henneaux:2015cda} and bridges the gap between the 
tensor formulation of \cite{Henneaux:2015cda} and the original 
spinor  formulation of \cite{Pope:1989vj}. 

\section{Conclusion}
\label{sec:conclusions}
Although the problem of classifying the possible systems of unfolded equations 
for conformal higher spin fields is solved, \it explicitly \rm writing down such a set
of differential equations may be cumbersome for technical reasons.
In this paper, we embedded
the $\so(1,2)$ field content needed \cite{Pope:1989vj} to describe a spin-$s$ conformal 
field in a compact set of $\gl(3,\R)$-valued one-forms, 
which can in turn be packed up in a single $\gl(5,\R)\,$-valued connection. 

We showed that, although the Schouten tensor is not a gauge-invariant quantity 
--- and is not contained in any $\sigma^-$ cohomology class for the conformal spin-$s$
theory --- there is a unique gauge (called the $B$-gauge) 
where our simple embedding \eqref{Schoutendef} realises it and complies with the transformation law and Bianchi identity of \cite{Henneaux:2015cda}. 
In this gauge and using our $\gl(3,\R)$-covariant reformulation of the Pope-Towsend
system \cite{Pope:1989vj},
we could prove in yet a different way the cohomological problems analysed in 
\cite{Henneaux:2015cda}. Another advantage of the Cartan-like 
presentation of the equations of motion is that the extension to 
(A)dS$_3$ backgrounds is direct, as it is simply encoded in the choice of 
background connection $\Omega\,$. 

Some advantages we see of our approach are the following:
(1) We could bridge the gap between the conformal off-shell tensor calculus 
studied in \cite{Henneaux:2015cda} and the spinor calculus introduced 
in \cite{Pope:1989vj} and used by many other authors, in particular 
in \cite{Nilsson:2013tva,Nilsson:2015pua} for recent investigations 
towards nonlinearities including matter couplings; 
(2) we could present a manifestly 3-dimensional off-shell conformal tensor 
calculus while at the same time avoiding the technicalities involved in dealing with  
$\so(2,3)$ projectors on two-row Young tableaux; 
(3) we wrote down a set of unfolded equations that, because they are devoid 
of any spinor technologies, are amenable to extensions to higher dimensions 
as they resort to one-form connections  
that mimic the Lopatin--Vasiliev connections \cite{Lopatin:1987hz}.

Finally, we believe that our reformulation of the conformal spin-$s$ equations 
\cite{Pope:1989vj} can be useful for a systematic investigation of 
the possible interactions among conformal spin-$s$ fields, in 3 and higher 
dimensions. We hope to come to this issue in the near future.

\section*{Acknowledgements}
N.B. thanks Ergin Sezgin for having suggested the problem, 
at the time the paper \cite{Boulanger:2014vya} was being finalised.
He also acknowledges stimulating discussions with Marc Henneaux 
at the Higher Spin workshop in Singapore.
We thank Andrea Campoleoni and Jordan Fran\c cois for useful discussions. 
T.B. is supported by a joint grant ``50/50'' Universit\'e
Fran\c{c}ois Rabelais Tours -- R\'egion Centre / UMONS.
The work of R.B. was supported by a PDR ``Gravity and extensions'' from 
the F.R.S.-FNRS (Belgium).







\providecommand{\href}[2]{#2}\begingroup\raggedright
\endgroup

\end{document}